\documentclass[article]{JHEP}

\usepackage{amssymb,amsmath,amsfonts}
\usepackage{nicefrac}
\usepackage{epsfig}

\relax
\renewcommand{\theequation}{\arabic{section}.\arabic{equation}}

\newcommand\fverb{\setbox\pippobox=\hbox\bgroup\verb}
\newcommand\fverbdo{\egroup\medskip\noindent%
                        \fbox{\unhbox\pippobox}\ }
\newcommand\fverbit{\egroup\item[\fbox{\unhbox\pippobox}]}
\newbox\pippobox

\def\be{\begin{equation}}
\def\ee{\end{equation}}

\def\FF{{\cal F}}
\def\GG{{\cal G}}

\def\II{{\cal I}}

\def\KK{{\cal K}}
\def\LL{{\cal L}}
\def\MM{{\cal M}}
\def\NN{{\cal N}}
\def\OO{{\cal O}}
\def\PP{{\cal P}}

\def\SS{{\cal S}}

\def\UU{{\cal U}}
\def\VV{{\cal V}}

\def\ZZ{{\cal Z}}

\def\d{{\partial}}
\def\llangle{{\langle \langle}}
\def\rrangle{{\rangle \rangle}}

\newcommand{\beq}{\begin{equation}}
\newcommand{\eeq}{\end{equation}}
\newcommand{\ba}{\begin{array}}
\newcommand{\ea}{\end{array}}
\newcommand{\bea}{\begin{eqnarray}}
\newcommand{\eea}{\end{eqnarray}}

\newcommand{\tr}{\mathop{{\rm Tr}}}

\def\hre#1#2{\href{http://arxiv.org/abs/#1/#2}{[ArXiv:#1/#2]}}

\def\hri#1#2{\href{http://arxiv.org/abs/#1}{[ArXiv:#1]#2}}

\numberwithin{equation}{section}

\title{(Multi)Matrix Models and Interacting Clones of Liouville Gravity}
\author{Elias Kiritsis$^{\spadesuit,\diamondsuit}$ and Vasilis Niarchos$^{\spadesuit}$
\\ \hspace*{\fill}
\\$^{\spadesuit}$Centre de Physique Th\'eorique, \'Ecole Polytechnique,
91128 Palaiseau, France
\\ Unit\'e mixte de Recherche 7644, CNRS\\ \hspace*{\fill}
\\$^\diamondsuit$Department of Physics, University of Crete, 71003 Heraklion, Greece
}

\preprint{CPHT-RR 030.0508}

\abstract{Large-$N$ matrix models coupled via multitrace operators are used to define,
via appropriate double-scaling limits, solvable models of interacting multi-string
theories. It is shown that although such theories are non-local at the world-sheet level
they have a simple description of the spacetime physics. Such theories share the main
characteristics of similarly coupled higher-dimensional CFTs. An interpretation has
been given in the past of similar continuum limits in terms of Liouville interactions
that violate the Seiberg bound. We provide a novel interpretation of this relation
which agrees with the current understanding of Liouville theory and analogous
observations in the AdS/CFT correspondence.}

\begin{document}

\section{Introduction}
\label{sec:intro}

\subsection{Setup and questions}

In the context of the AdS/CFT correspondence, it has been argued
\cite{kir,ack} that the holographic dual of the product of $k$ conformal
field theories (CFTs) in $d$ dimensions, deformed by a set
of multi-trace interactions that couple the CFTs together, is a theory of quantum
(multi)-gravity (or better a multi-string theory) on a union of $k$ $AdS_{d+1}$
spaces with a formally common boundary -- the boundary being isomorphic
to the space where the dual product CFT lives. Assuming that each CFT has
an independent gauge group $G_i$ and that there are no fields charged under
more than one group, multi-trace interactions are the only way to couple the
CFTs. We will assume in this paper that we are working in the large-$N$ limit.

From the quantum gravity point of view this is an interesting setup for
the following reasons:
\begin{itemize}
\item[$(i)$] This is a non-trivial example of an interacting multi-graviton theory
with a UV completion (the completion provided by the dual CFT).
$1/N$ effects generate a non-vanishing potential for the gravitons
making some of them massive. This effect appears as follows: on the
field theory side, the multi-trace interactions violate the conservation of the
energy-momentum tensor of each CFT, they retain, however, the conservation of
a total stress tensor. On the dual gravity side,
a linear combination of the $k$ original gravitons remains massless while
the rest of the gravitons obtain non-zero masses of order $1/N$.
One can think of this effect as a Higgs mechanism for gravity \cite{porrati}.

Coupling more than one gravitons together, or giving the gravitons a mass, have been
long standing theoretical problems \cite{fp,deser,vain,ags,insta,bachas,henneaux}
(see \cite{Rubakov:2008nh} for a recent review). The issue at stake is the possibility
of coupling non-trivially massless gravitons, or giving them a mass without rending the
theory UV sensitive. So far there are no-go theorems for non-trivial couplings among
massless gravitons \cite{henneaux}. On the other hand, most attempts to write an
effective low-energy action for massive graviton theories are plagued by serious
problems -- ghosts, instabilities and strong coupling problems.

This behavior is not only characteristic of theories where a graviton has a mass term
in the Lagrangian, but also more exotic cases where the massive graviton can be a
resonance. This happens when gravity is induced on branes, with the most
celebrated example being given by the DGP model \cite{dgp}.  Indeed it was
observed in \cite{ktt,dr} that this theory exhibits a similar non-decoupling behavior
as the standard Pauli-Fierz theory. It was subsequently shown that the theory
becomes strongly coupled at hierarchically low scales \cite{lpr,rub}. Moreover,
as has already been analyzed in detail in \cite{deser,insta} massive graviton theories
are generically unstable. Instabilities similar to those of the Pauli-Fierz theory also occur
in brane-induced gravity as reviewed in \cite{greg}.

In this respect, having an example with a UV complete, non-perturbative
formulation, where these issues can be  addressed, is important not only from an
academic desire to know if such theories exist, but also for potential phenomenological
applications. Infrared modifications of gravity of this type could be useful, for example,
in the resolution of the cosmological constant.
\item[$(ii)$] It is interesting to ask if and how standard properties of string theory
and gravity (at the classical and quantum level) are modified in a multi-verse
of interacting worlds. One can imagine new qualitative
features on the string theory worldsheet and new dynamics in spacetime with
potentially useful applications -- for instance, potential applications in
cosmology ($c.f.$ \cite{Damour:2002wu} for a search of cosmological solutions
in a multi-gravity theory).

The above AdS/CFT example offers a concrete arena to study
the possibility and structure of such effects. In the context of the AdS/CFT correspondence,
it has been argued \cite{nonlocal} that multi-trace deformations
on the boundary theory re-arrange the string perturbation theory in the bulk leading to
a non-local string theory (NLST) with non-local dynamics on the worldsheet.
One would like to know the precise rules of such dynamics. In particular, we
would like to know whether such theories really define a new universality class of string
theories. In this paper, we will have the chance to address some of these questions
in a set of low-dimensional examples, where string theory is solvable.
\end{itemize}

\subsection{Deformed matrix model products and non-critical NLSTs}

The example of product CFTs and string theory on a union of AdS spacetimes can
be generalized if we consider the product of more general quantum field theories
(QFTs). As long as each of these QFTs has a string theory dual, the dual of the product
theory will be a string theory on a union of spacetimes.

One of the first and best understood examples of
holographic duality in string theory is the duality between
the double scaling limit of large $N$ matrix models and two-dimensional quantum
gravity coupled to conformal matter with central charge $c\leq 1$, $i.e.$ $c\leq 1$
non-critical strings (for a review see \cite{Klebanov:1991qa,Ginsparg:1993is} and
references therein). This is a good context for some of the above questions, because
string theory in these examples is solvable at all orders in perturbation theory.
Unfortunately, this case will not allow us to address the issues raised in point $(i)$ above --
the spacetime theory is a theory of scalar fields, hence we cannot arrange for a
multi-gravity theory in this  context. It allows us, however, to address some of the
questions in point $(ii)$ in a precise manner. Higher dimensional AdS/CFT examples
that involve a multi-gravity theory and the related issues in point
$(i)$ will be discussed separately in a companion paper \cite{kirniar2}
(see, however, subsection \ref{subsec:ads} below and the discussion in
section \ref{discussion} for a summary of the main results).

The precise setup we want to consider in this paper is as follows. The ``boundary''
theory involves a product of $k$ large-$N$ matrix models $\prod_{i=1}^k \MM_i$.
The simultaneous standard double-scaling limit in each factor $\MM_i$ gives
the holographic description of a product of $c\leq 1$ non-critical string theories
$\prod_{i=1}^k \SS_i$, where each factor in this product is independent and
does not communicate with the rest. We want to deform the product
$\prod_{i=1}^k \MM_i$ by adding multi-trace interaction terms to the total
matrix model Lagrangian that couple different $\MM_i$'s together. The
non-trivial question is: under what circumstances can we find new
double-scaling limits with non-vanishing multi-trace interaction couplings?
Such limits would define holographically a multi-verse of interacting
$c\leq 1$ non-critical string theories.

In section \ref{sec:matrixmodels} we analyze the product of two Hermitian
one-matrix models deformed by a general double-trace deformation.
We will find that in this case a one-parameter family of double scaling limits
exists with a non-trivial coupling between the two matrix models provided that
the scaling properties of the single-trace operators that participate in the
deformation are the same and that we tune the double-trace parameters to a
special set of values. The free energy $\tilde F$ of the deformed theory is no
longer the direct sum of the free energies $F_1$ and $F_2$ of matrix models $\MM_1$
and $\MM_2$ respectively. It is the logarithm of a bilateral Laplace transform
\beq
\label{introaa}
\tilde F(\tilde t_+,t_-)=\log \int_{-\infty}^\infty dt_+ ~
e^{\tilde t_+ t_+ + F_1(\sin \theta~ t_++\cos\theta ~t_-)+
F_2(\cos \theta ~t_+ - \sin \theta~t_-)}
~,
\eeq
where $\theta \in [0,2\pi)$ is a free angular variable parametrizing a $U(1)$
subspace of the three-dimensional double-trace coupling space and $\tilde t_+, t_-$
are scaling parameters in the modified two-matrix model. This result
is a natural generalization of a similar formula that arises in double-trace
deformations of single Hermitian one-matrix models \cite{Klebanov:1994pv,
Klebanov:1994kv}. By holography, it gives the free energy of two $coupled$ minimal
string theories.

In section \ref{correlations} we find that
the genus expansion of correlation functions in the deformed theory
\eqref{introaa} boils down, when expressed in terms of correlation functions
in the undeformed product  $\MM_1 \otimes \MM_2$, to a diagrammatic
expansion where one has to sum over a series of terms that involve
surfaces with contact interactions between the world-sheets of theory 1 and 2.
This makes the world-sheet theory non-local. At tree-level, where \eqref{introaa}
reduces to a Legendre transform, all the contributing world-sheets are genus zero,
however, at any higher loop order $g\geq 1$ there is a series of contributions from
touching surfaces with different geni less or equal than $g$. This is a concrete
example of a non-local string theory expansion along the lines of \cite{nonlocal}.

In higher dimensional AdS/CFT systems, in a limit where string theory
in the bulk reduces to semi-classical gravity (this is the limit of large 't Hooft
coupling for a gauge theory on the boundary), there is, at tree-level, an alternative
reformulation of the gravity dual of multi-trace deformations as ordinary gravity with
mixed boundary conditions for the dual fields \cite{witten,berkooz,muck,minces,petkou}.

A similar reformulation of the tree-level theory appears to be possible in
the context of matrix models and non-critical strings. This is in effect an old observation
of Klebanov \cite{Klebanov:1994pv}, who pointed out that the string susceptibility
exponents and tree-level correlation functions in the double-trace deformed theory
can be interpreted in the dual string theory as a change of the Liouville dressing of the tachyon
condensate from the right branch to the wrong branch. At face value, this is a
peculiar statement in Liouville theory. It would seem to suggest that we have to change
Liouville theory in a drastic way by modifying the boundary conditions of the tachyon
field at infinity. This is at odds with our current understanding of Liouville theory, where
both branches (or equivalently both the standard and the dual cosmological constant
interactions) have to be present in order to account for the right structure of correlation
functions \cite{Dorn:1992at,Dorn:1994xn,Zamolodchikov:1995aa,Teschner:2001rv,
Teschner:2003en}. To the same effect, the only solution of the
Wheeler-DeWitt equation that is regular in the strong coupling region has weak
coupling asymptotics with both branches turned on \cite{Ginsparg:1993is}.
From this point of view, we can change the weak coupling asymptotics at the
expense of introducing a singularity in the strong coupling region. The singularity
could have a physical origin, $e.g.$ it could be attributed to a bunch of
ZZ branes \cite{Zamolodchikov:2001ah} localized in the strong coupling region.
It is unclear, however, why ZZ branes would suddenly make an appearance in
this context.

In section \ref{spacetime} we will propose a different interpretation
of the observations in \cite{Klebanov:1994pv}, which agrees
with the current understanding of Liouville theory and analogous observations in the
AdS/CFT correspondence. The basic point is this. In Liouville theory the cosmological
constant $\mu$ and the dual cosmological constant $\tilde \mu$ are both present
and have a fixed relation. The theory is not modified if we exchange at the same time
$\mu \leftrightarrow \tilde \mu$ and the branch of the Liouville interaction.
However, each of these transformations separately will give a different theory.
We propose that this theory is what the modified matrix model describes at tree-level.
In this modification Liouville theory continues to obey the usual rules. What
we modify is the definition of the scaling parameter or in other words the specifics of the
bulk/boundary dictionary. This is precisely what happens also in the AdS/CFT
correspondence when we talk about mixed boundary conditions.

Beyond tree-level the standard free energy evaluated as a function of $\tilde \mu$
does not agree with the exact results obtained from the Laplace transform \eqref{introaa}.
This implies that the modified theory in the bulk is truly in a new universality
class of string theories with a non-local worldsheet theory as envisioned in \cite{nonlocal}.
These points will be discussed further in section \ref{spacetime} (and appendix
\ref{app:susceptibility}).

There are several extensions of the modified product of two Hermitian one-matrix
models that we explore in the main text (section \ref{sec:matrixmodels} and appendices
\ref{app:MQM}, \ref{app:tri}). One is the extension to matrix quantum mechanics
and the dual bosonic $c=1$ string theory. Another is the possibility to couple
more than two non-critical string theories either by interactions that couple pairs
of string theories or by higher-order `vertices' that involve higher multi-trace interactions.
As an illustration, appendix \ref{app:tri} considers double scaling limits in
a product of three Hermitian one-matrix models coupled by a triple-trace
interaction.

\subsection{AdS/CFT and multi-gravity}
\label{subsec:ads}

A direct  analog of a double-scaled matrix model in higher dimensions is
a conformal field theory. In generic products of CFTs deformed by multi-trace
interactions, the multi-trace couplings break conformal invariance.
If $\Phi=\prod_{i=1}^n \OO_i$ is the perturbing operator in terms of a set of
single-trace operators $\OO_i$,  where $i$ labels the $i$-th CFT, then
there is an upper limit on $n$ in order for $\Phi$ to be perturbatively
relevant or marginal. This upper limit depends crucially on
the spacetime dimension $d$ through the unitarity bound. For scalar
operators $\OO_i$ the upper limit is 2 for $d\geq 6$, 3 for $d=5,4$,
5 for $d=3$ and infinity for $d=2$.

For concreteness, let us consider the case of two CFTs deformed by a
double-trace interaction. Analyzing the one-loop $\beta$-functions of single-trace
and multi-trace couplings in conformal perturbation theory
one recovers the following picture \cite{kirniar2}.

At tree-level, $i.e.$ to leading order in the $1/N$ expansion, there is a
one-parameter family of non-trivial fixed points provided that the single-trace
operators $\OO_i$ participating in the double-trace deformation have the same
scaling dimensions (or scaling dimensions with a sum equal to the spacetime
dimension). This is one of the features that we encounter also in the corresponding
matrix model analysis. In addition, one finds that the fixed points are repellors of the
renormalization group (RG) equations, and perturbing away from
them either drives the theory towards a fixed point, where the coupling
between the CFTs vanishes, or towards strong coupling outside the range
of validity of conformal perturbation theory.

The dual description of this system involves quantum gravity
on the union of two AdS spaces with mixed boundary conditions for
the scalar fields that are dual to $\OO_i$ \cite{kir,ack}. At tree-level, there are
two massless, non-interacting gravitons in this system and the RG running
of the double-trace couplings on the boundary is encoded subtly in the mixed
boundary conditions (it is not visible, in particular, as a radial running of the
background solution away from AdS).

$1/N$ effects modify this picture in an interesting way. On the gauge theory
side, $1/N$ effects shift the submanifold of fixed points and induce an RG running
of single-trace operators. On the gravity side, a non-trivial potential is generated
for the gravitons leading to an interacting bi-gravity theory in the general spirit
of Kogan and Damour \cite{Damour:2002ws}. A linear combination of the
gravitons remains massless, the orthogonal one obtains a mass of order $1/N$
\cite{kir,ack}. The quantum generated potential backreacts to the original AdS solution and
for generic bare values of the double-trace couplings the bulk solution is no
longer the union of two AdS spaces. It is rather a union of two spaces that are
radially deformed and interpolate between AdS spaces.

In cases, where we can trust the boundary conformal perturbation theory, the gauge
theory analysis predicts that the IR geometries are a union of AdS spaces with trivial
coupling, hence in this region of spacetime there are again two massless, non-interacting
gravitons. In a sense, the theory is dynamically washing away the mass of the graviton.
The only way to retain a non-trivial bi-gravity theory everywhere in spacetime
is to fine-tune the bare values of the double-trace couplings on a special
one-dimensional submanifold of fixed points. The general lesson seems to
be that massive gravity theories can in principle exist in a non-perturbative,
well-defined quantum gravity context, but are not generic and require a high
degree of fine-tuning. For a more detailed discussion
of the general AdS/CFT case we refer the reader to \cite{kirniar2}.

In the rest of this paper, we will deal with a solvable toy model of this general
setup that involves matrix models and non-critical strings. The analogies between
the matrix model and AdS/CFT setups will be summarized in the concluding section
\ref{discussion}.

\section{Matrix models and a multiverse of $c\leq 1$ string theories}
\label{sec:matrixmodels}

Multi-trace deformations in the context of matrix models and their implications
for Liouville gravity were discussed originally in
\cite{Das:1989fq,Klebanov:1994pv,Klebanov:1994kv}. In
this section, we will consider the effect of multi-trace deformations in a product of
matrix models. Extending the analysis of \cite{Klebanov:1994kv}, we are
looking for new double scaling limits that define holographically an interacting
union of $c\leq 1$ non-critical string theories. As a first concrete illustration,
we will discuss the case of two large-$N$ Hermitian matrix models
coupled by a double-trace deformation. Possible generalizations will be
discussed in the last subsection \ref{subsec:triple} and appendix \ref{app:tri}.

\subsection{Hermitian matrix models and minimal string theories}
\label{subsec:mms}

Let us begin by recalling some basic facts about multi-critical
Hermitian one-matrix models and their double scaling limits. The partition
function of the $k$-th multicritical one-matrix model is defined by the matrix
integral
\beq
\label{mmaaa}
\ZZ_k=\int d\Phi ~ e^{-N\left[ \tr V_k(\Phi)+(c_2+\lambda) \tr \Phi^4 \right]}
~,
\eeq
with
\beq
\label{mmaab}
V_k(\Phi)=\sum_{i=1}^k (-)^{i+1} c_i  \Phi^{2i}
\eeq
and $c_i$ a set of known constants \cite{matrixmodels}. In \eqref{mmaaa}
we have chosen to consider a deformation of the potential by a term
proportional to the single-trace operator $\tr \Phi^4$. This model has
a double scaling limit, where $N \to \infty$ and $c_2+\lambda \to 0$
with the scaling variable $t \sim (c_2+\lambda) N^{2k/(2k+1)}$ kept fixed.
In this limit, the partition function $\FF_k=\log \ZZ_k$ becomes a function
of the scaling variable $t$ and the matrix model provides the holographic
formulation of the minimal $(2,2k-1)$ bosonic string \cite{matrixmodels}.

We will consider a two-matrix model, which arises from
the direct product of a $k$-th and a $p$-th multi-critical matrix
model ${\cal M}_{2,2k-1}\otimes {\cal M}_{2,2p-1}$ by a double-trace
deformation. The standard double scaling limit of this theory at zero
double-trace coupling describes the decoupled union of a $(2,2k-1)$ and a
$(2,2p-1)$ minimal string.

\subsubsection{A prototype: deforming the ${\cal M}_{2,3}\otimes {\cal M}_{2,3}$ product}

The theory we want to solve is a double-trace deformation of the
${\cal M}_{2,3}\otimes {\cal M}_{2,3}$ two-matrix model with partition function
\beq
\label{mmaac}
\ZZ=\int D\Phi_1 D\Phi_2 e^{-N_1 \tr \big[ \frac{1}{2}\Phi_1^2+\lambda_1 \Phi_1^4\big]
-N_2\big[\frac{1}{2}\Phi_2^2+\lambda_2 \Phi_2^4 \big]
-\big[ g_{11} (\tr \Phi_1^4)^2+g_{22} (\tr \Phi_2^4)^2+2 g_{12} \tr \Phi_1^4 \tr \Phi_2^4
\big]}
~.
\eeq
The double-trace deformation involves the operators $\tr \Phi_1^4$, $\tr \Phi_2^4$.
It is $g_{12}$ that couples the two separate matrix models, but $g_{11}$ and $g_{22}$
will play an important r\^ole when we take double scaling limits.
The ranks of the matrices $\Phi_1,\Phi_2$ are $N_1,N_2$ respectively.
In this paper we use the notation
\beq
\label{mmranks}
N_1 \equiv  N~, ~ ~ N_2\equiv \nu N
~.
\eeq
In the large $N$ limit both $N_1$ and $N_2$ will
be scaled to infinity; the ratio $\nu$ will be kept fixed and
will be treated as an extra parameter of the system. The single-trace
and double-trace coupling constants are such that the overall Lagrangian scales
like $N^2$.

It will be convenient to re-express the double-trace deformation as
a sum of two squares
\bea
\label{mmaad}
&&g_{11} (\tr \Phi_1^4)^2+g_{22} (\tr \Phi_2^4)^2+2 g_{12} \tr \Phi_1^4 \tr \Phi_2^4
=\nonumber\\
&&=r_1 \left( \cos\theta \tr \Phi_1^4 + \sin\theta \tr \Phi_2^4\right)^2+
r_2(-\sin \theta \tr \Phi_1^4+\cos\theta \tr \Phi_2^4)^2
~.
\eea
Setting for quick reference
\beq
\label{mmaae}
C\equiv \cos\theta~, ~ ~ S\equiv \sin\theta
~
\eeq
we deduce the double-trace coupling relations
\beq
\label{mmaaf}
g_{11}=r_1 C^2+r_2 S^2~, ~ ~
g_{22}=r_1 S^2+r_2 C^2~, ~ ~
g_{12}=(r_1-r_2) SC
~.
\eeq
With these definitions
\beq
\label{mmaag}
\det g=g_{11}g_{22}-g_{12}^2=r_1 r_2~, ~ ~
\tr g= g_{11}+g_{22}=r_1+r_2
~.
\eeq

We shall distinguish between two different cases: $(i)$ $r_1 r_2 \neq 0$ and
$(ii)$ $r_1 r_2=0$. In the first case, the double-trace deformation consists of
two quadratic terms. In the second case, there is a single quadratic term
or no deformation at all.

\subsubsection*{$\bullet$ Case $(i)$: $r_1 r_2 \neq 0$}

Let us define the free energy $\hat \FF$ of the single-trace theory as
\beq
\label{mmaai}
e^{\hat \FF[\lambda,N^2]}=
\int D\Phi~ e^{-N \tr \big[ \frac{1}{2}\Phi^2+\lambda \Phi^4\big]}
~.
\eeq
A common trick to deal with the double-trace deformations
is to use the identity
\beq
\label{mmaaj}
e^{g\OO^2}=\frac{N}{\sqrt{4\pi g}}
\int_{-\infty}^\infty dy ~ e^{-\frac{N^2 y^2}{4g}} e^{Ny \OO}
~.
\eeq
In the case of \eqref{mmaac}, this trick allows us to recast the
partition function as a double integral over
single-trace parameters with a specific Gaussian weight
\beq
\label{mmaak}
\ZZ=\frac{N_1N_2}{4\pi \nu\sqrt {r_1r_2}} \int_{-\infty}^\infty dy_1 dy_2~
e^{\hat \FF\big[ \lambda_1-Cy_1+Sy_2,N_1^2\big]+
\hat \FF \big[\lambda_2-\frac{1}{\nu}(Sy_1+C y_2),N_2^2\big]}
e^{\frac{N^2}{4}\big[ \frac{y_1^2}{r_1}+\frac{y_2^2}{r_2}\big]}
~.
\eeq

It is now possible to rewrite $\ZZ$ in a more explicit form thanks to
the well-known solution of the one-matrix model \cite{Brezin:1977sv,matrixmodels}.
For the $k=2$ multi-critical matrix model, in particular, the free energy reads
\beq
\label{mmaal}
\hat \FF [\lambda,N^2]=N^2\left(-a_1 x +\frac{1}{2} a_2 x^2\right)+F(x,N^2)
~,
\eeq
where
\bea
\label{mmaam}
F(x,N^2)=&&N^2 \left( -\frac{2}{5} a_3 x^{5/2}+...\right)+
N^0\left(-\frac{1}{24} \log x+...\right)+\nonumber\\
&&+N^{-2}\left(a_4 x^{-5/2}+...\right)+
\OO(N^{-4})
~,
\eea
with
\beq
\label{mmaan}
x=c_2+\lambda
~, ~~
a_1=4~, ~ ~ a_2=576~,  ~ ~ a_3=6144\sqrt 3~, ~ ~ c_2=\frac{1}{48}
~, ~~ \cdots
\eeq
The omitted terms inside each parenthesis in the expression of $F$
\eqref{mmaam} will be subleading in the double-scaling limit  at each
order in $N$ and will not play a r\^ole in our discussion.

Our case involves the free energy of two decoupled one-matrix models
with parameters $x_1$, $x_2$. These parameters, which are to be integrated
over in \eqref{mmaak}, are related to the variables $y_1,y_2$ by the linear
transformation
\begin{subequations}
\bea
\label{mmaao}
y_1&=&\Delta_1-Cx_1-Sx_2~, ~ ~ \Delta_1=C\Lambda_1+\nu S\Lambda_2
\\
\label{mmaaoo}
y_2&=&\Delta_2+Sx_1-Cx_2~, ~ ~ \Delta_2=-S\Lambda_1+\nu C\Lambda_2
~,
\eea
\end{subequations}
where we define
\beq
\label{mmaaoa}
\Lambda_1=c_2+\lambda_1~, ~ ~ \Lambda_2=c_2+\lambda_2
~.
\eeq

Inserting this information into \eqref{mmaak} and dropping an unimportant
overall constant factor we obtain the partition function
\bea
\label{mmaap}
\ZZ&=&N^2 \int_{-\infty}^\infty dx_1 dx_2~
e^{\frac{N^2}{2}\big[ \big( -2a_1-\frac{\Delta_1C}{r_1}+\frac{\Delta_2 S}{r_2}\big)x_1
+\big( -2a_1\nu-\frac{\Delta_1 S}{r_1}-\frac{\Delta_2 C}{r_2}\big) x_2\big]}\times
\\
&&e^{\frac{N^2}{4}\big[ \big( 2a_2+\frac{C^2}{r_1}+\frac{S^2}{r_2}
\big)x_1^2+\big(2a_2+\frac{S^2}{r_1}+\frac{C^2}{r_2}\big)x_2^2+2CS
\big(\frac{1}{r_1}-\frac{1}{r_2}\big)x_1x_2\big]}
~e^{F(x_1,N_1^2)+F(x_2/\nu,N_2^2)}
~.\nonumber
\eea
To diagonalize the quadratic term in the exponent of the integrand
we rotate from $(x_1,x_2)$ to $(x_+,x_-)$:
\beq
\label{mmaaq}
x_1=U^+_1 x_+ +U^-_1 x_-~, ~ ~
x_2=U^+_2 x_+ +U^-_2 x_-
~,
\eeq
where
\beq
\label{mmaar}
U_1^+=S~, ~ ~ U_1^-=C
~,~~
U_2^+=C~,  ~ ~ U_2^-=-S
~.
\eeq
In terms of the rotated variables
\bea
\label{mmaas}
\label{bana}
\ZZ=N^2 \int_{-\infty}^\infty dx_+ dx_- &&
e^{\frac{N^2}{2}(E^+ x_+ +E^- x_-)} e^{-N^2m_+^2 x_+^2-N^2m_-^2 x_-^2}
\times
\nonumber\\
&&e^{F(U_1^+ x_++ U_1^- x_-, N_1^2)+F((U_2^+ x_+ +U_2^- x_-)/\nu,N_2^2)}
~,
\eea
where
\begin{subequations}
\beq
\label{mmaat}
E^{\pm}=\left( -2a_1-\frac{\Delta_1 C}{r_1}+\frac{\Delta_2 S}{r_2}\right) U_1^\pm
-\left( 2a_1\nu+\frac{\Delta_1 S}{r_1}+\frac{\Delta_2 C}{r_2}\right)U_2^\pm
~,
\eeq
\beq
\label{mmaau}
m_\pm^2=-\frac{1}{8} \left( \frac{1}{r_1}+\frac{1}{r_2}+4a_2\pm \frac{|r_1-r_2|}{r_1r_2}\right)
~.
\eeq
\end{subequations}

Depending on the sign of the saddle point mass squared parameters $m_\pm^2$
we can define a variety of double scaling limits. We will distinguish between the
following alternatives: $(a)$ both $m_\pm^2$ are positive, $(b)$ one mass$^2$ is positive,
the other is zero, $(c)$ both masses are zero, or $(d)$ at least one mass is
tachyonic.

$(a)$ occurs if and only if
\beq
\label{mmaba}
\frac{1}{r_1}<-2a_2~, ~ ~ \frac{1}{r_2}< -2a_2
~.
\eeq
In this case, the only sensible double scaling limit that we can take requires
\beq
\label{mmabb}
N \to \infty~, ~ ~ E^{\pm} \to 0~, ~ ~ {\rm so~ that} ~ ~
\frac{E^\pm}{4m_\pm^2} N^{4/5} a_3^{2/5}=t_\pm ~ ~ {\rm is~ kept~ fixed}
~.
\eeq
After standard manipulations (see \cite{Klebanov:1994kv}), we deduce
the partition function
\beq
\label{mmabc}
\ZZ=e^{F(U^+_1 t_+ +U^-_1 t_-)+F(\nu^{-1/5}(U^+_2 t_++U^-_2 t_-))}
=e^{F(t_1)+F(t_2)}=\ZZ(t_1)\ZZ(t_2)
~
\eeq
with the obvious definition of the scaling parameters $t_1$ and $t_2$.
The resulting theory is the product of two undeformed, decoupled matrix
models, which describe the union of two decoupled $(2,3)$ minimal strings.
The double scaling limit has driven the theory back to the undeformed
point where the effects of the double-trace deformation are washed away.
A similar effect in the case of a single matrix model was observed in
\cite{Klebanov:1994kv}.

This behavior is reminiscent of what happens with an irrelevant perturbation in
a higher dimensional quantum field theory when we follow the renormalization
group flow towards the infrared. In our matrix model example, going towards
the infrared is achieved by the double scaling limit which zooms around a critical
point. When the condition \eqref{mmaba} holds, the effects of the double-trace
deformation \eqref{mmaad} disappear in the double scaling limit and the perturbation
behaves as an irrelevant operator. In fact, this is more than an analogy: in higher
dimensions there is a range of parameters where double-trace perturbations
generalizing \eqref{mmaad} are indeed irrelevant in the usual RG sense \cite{kirniar2}.

Case $(b)$ is more interesting. Now $m_+^2=0$ and $m_-^2>0$,
which is equivalent to
\beq
\label{mmabd}
\left(\frac{1}{r_1}+2a_2\right)\left( \frac{1}{r_2}+2a_2\right)=0
~.
\eeq
This condition is satisfied by a two-parameter family of deformations.
Notice that under \eqref{mmabd} it is impossible to achieve a non-vanishing
$g_{12}$ without turning on at the same time the other two couplings
$g_{11},g_{22}$.

To see what happens to the free energy, we set $m_+^2=0$ in
\eqref{mmaas}, and take the double scaling limit
\bea
\label{mmabe}
&&N\to \infty~, ~ ~ E^\pm \to 0~, ~ ~ t_+=x_+ N^{4/5} a_3^{2/5} ~ {\rm fixed}~, ~ ~
\nonumber\\
&&t_-=\frac{E^-}{4m_-^2} N^{4/5} a_3^{2/5}~ {\rm fixed}~, ~ ~
\tilde t_+=\frac{E^+}{2} N^{6/5} a_3^{-2/5}~ {\rm fixed}
~.
\eea
After standard manipulations the double scaled partition sum becomes
\beq
\label{mmabf}
\ZZ(\tilde t_+,t_-)=\int_{-\infty}^\infty dt_+ ~ e^{\tilde t_+ t_+
+F(U_1^+ t_+ +U^-_1t_-)+F(\nu^{-1/5}(U^+_2 t_+ +U^-_2 t_-))}
~.
\eeq
$\ZZ$ depends now on the double scaling parameters $\tilde t_+, t_-$
and parametrically on $\theta$, which is one of the deformation
parameters \eqref{mmaad}.

Eq.\ \eqref{mmabf} is an interesting exact formula that deserves a few comments.
First, we observe that the old scaling parameter $t_+$ has been transmuted
to a new scaling parameter $\tilde t_+$. $t_+$ requires scaling with $N^{4/5}$,
whereas $\tilde t_+$ requires scaling with $N^{6/5}$. A similar change of critical behavior
occurs in a single hermitian matrix model deformed by a double-trace operator
as was pointed out originally in \cite{Klebanov:1994pv,Klebanov:1994kv}.
In the single matrix model case with a double-trace deformation of the form
$g(\tr \Phi^4)^2$, the new critical behavior occurs when $g=-\frac{1}{2a_2}$
and exhibits the partition function
\beq
\label{mmsingle}
\ZZ(\tilde t)=\int_{-\infty}^\infty dt ~ e^{\tilde t t+F(t)}
~.
\eeq
Eq.\ \eqref{mmabf} generalizes this effect to the two-matrix example \eqref{mmaac}.

Both \eqref{mmabf} and \eqref{mmsingle} are bilateral Laplace transforms
of the partition sum of the original undeformed theory. For non-vanishing
$g_{12}$ \eqref{mmabf} is $not$ anymore the partition function of the
decoupled product of two one-matrix models. The dual minimal string
interpretation of this formula will be discussed in section \ref{spacetime}.
For $g_{12}=0$ we expect to recover the partition function of a decoupled product of
theories. This partition function is factorizable with one factor being the double scaled
partition function of an undeformed one-matrix model and the other being
the Laplace transformed partition function as in \eqref{mmsingle}.
Indeed, we can check that when $g_{12}=0$
\beq
\label{mmabg}
\ZZ=e^{F(\nu^{-1/5}t_-)} \int_{-\infty}^\infty dt_+ e^{\tilde t_+ t_+ +F(t_+)}
~.
\eeq

The non-perturbative meaning of eqs.\ \eqref{mmabf} -- \eqref{mmabg}
is not completely clear. For instance, there are well-known problems in defining
the one-matrix free energies $F(t)$ non-perturbatively. Presumably, we can avoid
this problem by looking at a different class of matrix models, which are non-perturbatively
well-defined (see $e.g.$ \cite{Klebanov:2003wg} and references therein). Even then,
however, we have to check the convergence of the integral appearing in the Laplace
transform or whatever generalizes the Laplace transform. Despite these important
potential issues, eqs.\ \eqref{mmabf} -- \eqref{mmabg} work well at any order in perturbation
theory, which is what we will focus on in this paper.

As a final comment on this case, we point out that there is an interesting analog
of the critical behavior captured by eqs.\ \eqref{mmabf}, \eqref{mmabg} in higher
dimensional AdS/CFT examples \cite{kirniar2}. In conformal perturbation theory
one finds again a one-parameter family of interacting fixed points with non-vanishing
double-trace couplings, which translate in the AdS/CFT correspondence to a system
of coupled string theories on a union of AdS spaces.

When both masses in \eqref{mmaau} are taken to be zero (this is case $(c)$
above) we obtain
\beq
\label{mmabi}
g_{12}=0~{\rm and}~ g_{11}=g_{22}=-\frac{1}{2a_2}~ {\rm or} ~0
~.
\eeq
With $g_{11}=g_{22}=-\frac{1}{2a_2}$ we are dealing with the
decoupled product of two one-matrix models tuned to their individual
double-trace deformed critical points that were analyzed in \cite{Klebanov:1994kv}
(see eq.\ \eqref{mmsingle}).

Finally, when at least one of the masses $m_\pm$ is tachyonic, the theory goes
into a branched polymer phase where we cannot define anymore a double
scaling limit describing a sum over continuous surfaces \cite{Klebanov:1994kv}.

\subsubsection*{$\bullet$ Case $(ii)$: $r_1 r_2=0$}

Similar manipulations can be performed when the determinant
of the matrix of double-trace parameters $g_{ij}$ vanishes, $i.e.$ when
$r_1r_2=0$. In this case, there is only one integration over single-trace
deformation parameters in \eqref{mmaak} and one can obtain again
different critical behaviors depending on the sign of the saddle point mass
squared $m^2$. In the stability region $m^2>0$, we recover \eqref{mmabc},
the decoupled product of two one-matrix models. When the mass $m$ is zero,
which occurs for a one-parameter family of double-trace deformations
parametrized by the angle $\theta$, we recover the interacting
product of one-matrix models \eqref{mmabf}. When the mass is tachyonic
the theory goes into a branched polymer phase.

\subsubsection{Comments on the general ${\cal M}_{2,2k-1}
\otimes {\cal M}_{2,2p-1}$ product}

Much of what we said above about the ${\cal M}_{2,3}\otimes {\cal M}_{2,3}$
product goes through unchanged to the more general
${\cal M}_{2,2k-1}\otimes {\cal M}_{2,2p-1}$ product of one-matrix
models, although some important changes in the conclusion occur when
$k\neq p$. In the general case, the double-trace deformed partition function reads
\bea
\label{mmabj}
\ZZ=\int D\Phi_1 D\Phi_2&&
e^{-N\big[ \tr V_k (\Phi_1)+(c_2+\lambda_1)\tr \Phi_1^4
+\tr V_p(\Phi_2)+(c_2+\lambda_2)\tr \Phi_2^4\big]}  \times
\nonumber\\
&&e^{-\big[ g_{11}(\tr \Phi_1^4)^2+g_{22}(\tr \Phi_2^4)^2+
2g_{12}\tr \Phi_1^4 \tr \Phi_2^4\big]}
~
\eea
with the potential $V_k$ still defined as in \eqref{mmaab}.

We recall that the free energy of the $k$-th multicritical matrix model is
\beq
\label{mmabk}
\hat \FF [\lambda]=N^2\left(-a_1 x+\frac{1}{2}a_2 x^2\right)+F_k(x,N^2)
~,
\eeq
where as before $x=c_2+\lambda$, but now more generally
\beq
\label{mmabl}
F_k(x,N^2)=-\frac{k}{2k+1}a_3 N^2 x^{(2k+1)/k}+...
\eeq
Because of the $k$ dependence of the non-singular part of the free energy
$F_k(x,N^2)$, the double scaling limit in the single-trace theory is
\beq
\label{mmabm}
N \to \infty~,~ ~ x \to 0~,~ ~ t=x N^{2k/(2k+1)} ~ {\rm fixed}
~.
\eeq
On the other hand, since the singular part of the free energy in \eqref{mmabk}
is $k$-independent, all the manipulations regarding this part
go through unchanged as in the $\MM_{2,3}\otimes \MM_{2,3}$ case.
In particular, the criteria for stability or instability remain the same.

Important differences occur when we take double scaling limits.
There are no substantial changes when $k=p$, in which case the
expressions we derived in the $\MM_{2,3} \otimes \MM_{2,3}$ case remain true with
the obvious modifications of double scaling limits.
On the other hand, when $k \neq p$ we find that
that there is no sensible double scaling limit that leads to the analog of
eq.\ \eqref{mmabf}. The reason for the absence of a sensible double scaling
limit is the difference between the scaling behaviors \eqref{mmabm} for $k\neq p$.
We observe a similar effect in conformal perturbation theory in higher dimensional
QFT examples \cite{kirniar2}. New fixed points with non-zero double-trace
coupling $g_{12}$ exist only when the single-trace operators $\OO_1,\OO_2$
participating in the double-trace deformation have the same scaling dimension.

\subsubsection{More general deformations of the
${\cal M}_{2,2k-1}\otimes {\cal M}_{2,2k-1}$ product}

For a single $k$-th matrix model new double scaling limits
can be defined by adding double-trace deformations $g \OO^2$ with
a general scaling operator $\OO$ after the necessary fine-tuning of
$g$ \cite{Klebanov:1994kv}. In that case, one obtains the free energy
\beq
\label{mmada}
\tilde \FF(t,\tilde t_O)= \log \int_{-\infty}^\infty dt_{\OO}~
e^{t_O \tilde t_O+\FF(t,t_O)}
~.
\eeq
This example can be generalized to
\beq
\label{mmadb}
\tilde \FF(\{ \tilde t \}, \{ T \})= \log \prod_{i=1}^n \int_{-\infty}^\infty
dt_i~ e^{\sum_{j=1}^n t_j \tilde t_j + \FF(\{ t \},\{ T \})}
~,
\eeq
where $\{T\}$ is the set of coupling constants that remain unintegrated.

Accordingly, for a product of $k$-th matrix models new double scaling limits
can be defined by tuning a set of multi-trace parameters. Consider, for instance,
two $k$-th one-matrix models deformed by the double trace operator
$g_{11} \OO_1^2+g_{22} \OO_2^2+2g_{12} \OO_1 \OO_2$, where
$\OO_1, \OO_2$ are respectively scaling operators in the matrix models $1$ and
$2$. We expect new double scaling limits leading to the free energies
\beq
\label{mmadc}
\tilde \FF(t_1,t_2,\tilde \tau_+, \tau_-)=\log \int_{-\infty}^\infty d\tau_+ ~
e^{\tau_+ \tilde \tau_+ +\FF_k(t_1,U_1^+ \tau_+ +U_1^- \tau_-)+
\FF_k(t_2,U_2^+\tau_+ +U_2^- \tau_-)}
~,
\eeq
where $t_1,t_2$ are single-trace couplings for the lowest dimension operators.
Further generalizations can be envisioned by introducing more couplings.

\subsection{Matrix quantum mechanics and $c=1$ string theories}
\label{subsec:MQM}

Analogous statements can be made for a pair
of matrix quantum mechanics theories coupled by a double-trace deformation.
In appendix \ref{app:MQM} we consider in some detail double-trace deformations
involving the cubic single-trace operator $\tr \Phi^3$. The partition function of
the theory we analyze there is
\bea
\label{qmaa}
\ZZ&=&\int D\Phi_1(t) D\Phi_2(t) ~ e^{-N \int_0^{2\pi R}dt
[\tr (\frac{1}{2} \dot\Phi_1^2+\frac{1}{2}\Phi_1^2-\lambda_1 \Phi_1^3)
+\tr (\frac{1}{2} \dot \Phi_2^2+\frac{1}{2}\Phi_2^2-\lambda_2 \Phi_2^3)]}\times
\nonumber\\
&&e^{-\int_0^{2\pi R}dt [g_{11}(\tr \Phi_1^3)^2+2g_{12} \tr \Phi_1^3 \tr \Phi_2^3+
g_{22}(\tr \Phi_2^3)^2]}
~.
\eea
The matrix model lives on a compact one-dimensional spacetime with radius $R$.
Besides this extra feature (that can be dealt with a Fourier transformation) most
of the elements of the analysis of the previous subsections go through also
in this case. The possible double scaling limits and the resulting expressions
are analogous to the zero-dimensional matrix model case above, so we will omit
a detailed discussion here. The interested reader can find a detailed analysis
of the partition function \eqref{qmaa} in appendix \ref{app:MQM}.

\subsection{Triple intersections and beyond}
\label{subsec:triple}

So far we have restricted our attention to pairs of matrix models.
In a suitable double scaling limit, these models provide the holographic
description of a product string theory with two components. There is a subtle
communication between the individual components of this product
induced by the double-trace deformation of the matrix model dual.
There are various possible generalizations of this picture.

We could consider, for instance, $n$ one-matrix models coupled together by
a general multi-trace interaction. We can imagine a large variety of possibilities.
One of them is to take $n$ theories coupled two-by-two via double-trace
deformations. The overall theory is an $n$-matrix model with an action of the form
\beq
\label{traa}
S=\sum_{i=1}^n S_i+\sum_{i,j=1}^n g_{ij} \OO_i \OO_j
~,
\eeq
where $S_i$ are the single-trace actions and $\OO_i$ are single-trace
operators, $e.g.$ $\OO_i=\tr \Phi_i^4$. Recasting $g_{ij}$ in terms of $n$
eigenvalues and an $O(n)$ matrix parametrized by $n$ orthogonal unit
vectors $\vec v^a$, we can write $S$ as
\beq
\label{trab}
S=\sum_{i=1}^n S_i+\sum_{a=1}^n r_a\left( \sum_{i=1}^n \vec v_i^a \OO_i \right)^2
~.
\eeq
The corresponding partition function can be analyzed as above
and allows for an obvious extension of the double scaling limits presented in previous
subsections provided that the operators $\OO_i$ have compatible scaling behaviors.

Other possibilities include higher multi-trace interactions with three or more
single-trace components. This allows for a more intricate set of double scaling
limits. As an illustration, in appendix \ref{app:tri} we analyze the triple intersection of three
2nd multicritical matrix models with the action
\beq
\label{trac}
S=\sum_{i=1}^n S_i+\frac{1}{N} \sum_{i,j,k=1}^3 g_{ijk} (\tr \Phi_i^4)
(\tr \Phi_j^4)(\tr \Phi_k^4)
~.
\eeq
For example, in the special case, where $g_{123}$ is the only non-vanishing
triple-trace coupling, we can find a double scaling limit with partition function
\beq
\label{trad}
\ZZ(\tilde t^1, \tilde t^2,t_3)= \int_{-\infty}^\infty dt_1 dt_2 ~
e^{\tilde t^1t_1+ \tilde t^2 t_2+ \sum_{i=1}^3 F(\UU^1_i t_1+\UU_i^2 t_2+\UU_i^3 t_3)}
~.
\eeq
$\UU^j_i$ are constants that can be determined. Once again we obtain a modified
partition function related to the undeformed one by a Laplace transform.
A more intricate pattern of double scaling limits is expected for quartic and higher
multi-trace deformations.

One qualitatively new feature in the triple-trace example, compared
to the double-trace case, is the possibility to find a non-trivial fixed
point with $g_{123}\neq 0$ and no other multi-trace couplings turned on.
This was not possible in the case of double-trace deformations, where at
points with $g_{11}=g_{22}=0, g_{12}\neq 0$ at least one of the saddle point
masses $m_\pm$ was tachyonic.

In higher dimensional QFTs unitarity and the requirement that the
$n$-trace interaction is a relevant or marginal operator puts an upper
bound on $n$. For example, in $d=4$ dimensions the bound is $n=3$.
The bound is absent only in two or lower dimensions. The general structure of the
one-loop $\beta$-functions for multi-trace interactions in $d$ dimensions will
be discussed in \cite{kirniar2}.

\section{Correlation functions in coupled minimal string theories}
\label{correlations}

In this section we will discuss scattering amplitudes
in a theory of coupled minimal strings using the matrix model
definition \eqref{mmabf}. We will illustrate the main points by recasting
the correlation functions of the deformed product theory as a diagrammatic
expansion in terms of correlation functions in the undeformed theory.
As an example, we will consider two-point functions at tree-level and one-loop.
The analysis in this section is a simple extension of the corresponding study
of correlation functions in a single modified minimal string in \cite{Barbon:1995dx}.
It will provide a concrete illustration of the non-local string theory structure
underlying the dynamics of the string theories of interest and will be a good guide
for the higher dimensional AdS/CFT examples discussed in the introduction,
where it is much harder to explore this structure explicitly beyond tree-level.

To be definite, let us focus on one of the cases discussed in section
\ref{sec:matrixmodels}: two $(2,2k-1)$ minimal strings coupled on the matrix
model side by a double-trace deformation. The partition function of the modified
theory is given by the Laplace transform
\beq
\label{appDaa}
e^{\tilde F(\tilde t_+,t_-;t_1,t_2)}=\int_{-\infty}^\infty dt_+ ~
e^{\lambda^{-1}\tilde t_+ t_+ +F_1(St_++Ct_-,t_1)+F_2(Ct_+-St_-,t_2)}
~.
\eeq
$t_1,t_2$ are single-trace couplings for the scaling operators $\OO_1$, $\OO_2$
of theory 1 and 2 respectively, whose correlations functions we will compute.
These operators do not partake in the double-trace deformation that couples
the two minimal strings. The constant $\lambda$ has been inserted in
\eqref{appDaa} to help us keep track of the genus expansion of the free
energies.

We can obtain the all-genus diagrammatic expansion of correlation functions
in the deformed theory by expanding the integral expression \eqref{appDaa}
around the saddle point value of $t_+$, which we will call $t_s$. $t_s$ is
determined by the following equation
\beq
\label{appDab}
\tilde t_+ + S\langle \PP_1 \rangle |_{St_s+Ct_-}+C\langle \PP_2 \rangle |_{Ct_s-St_-}=0
~,
\eeq
where the subindex next to the one-point functions denotes the point at
which the correlation functions are evaluated. In what follows, we will keep
this index implicit. $\PP_1$ and $\PP_2$ are operators in theory 1 and 2
respectively, whose single-trace couplings $St_++Ct_-$, $Ct_+-St_-$ appear
inside the integral in \eqref{appDaa}. In \eqref{appDab} we used the $n=1$
version of the identities
\beq
\label{appDac}
\lambda^{-1}\langle \PP_i^n \rangle |_{t} = \d_t^n F_i(t)~, ~ ~ i=1,2
~.
\eeq

The one-point functions in \eqref{appDab} receive contributions at all geni.
Since we are interested in a genus expansion (in other words, an expansion
in powers of $\lambda$) of $\tilde F$, it will be useful to define our point of
expansion by the tree-level version of \eqref{appDab}
\beq
\label{appDad}
\tilde t_++S\langle \PP_1 \rangle_0 +C\langle \PP_2 \rangle_0=0
~.
\eeq
The subindex in correlation functions will denote from now on the genus.

Setting $t_+=t_s+t$ for the integration variable in \eqref{appDaa}, we expand
around $t_s$ to obtain the free energy expression
\bea
\label{appDae}
&&\tilde F(\tilde t_+,t_-;t_1,t_2)=\lambda^{-1} \tilde t_+ t_s + F_1(St_s+C t_-,t_1)+
F_2(Ct_s-St_-,t_2)+
\\
&&+\log \int_{-\infty}^\infty dt~ \exp\left[ t \sum_{g=1}^\infty \lambda^{g-1}
\left(S \langle \PP_1\rangle_g+C\langle \PP_2 \rangle_g \right)+
\sum_{n =2}^\infty \sum_{g=0}^\infty \frac{1}{n!} t^n \lambda^{g-1}
\left( S^n \langle \PP_1^n\rangle_g +C^n \langle \PP_2^n\rangle_g\right)
\right]
\nonumber \, .
\eea
For example, from this expression we read off the tree-level and one-loop free energies
\begin{subequations}
\beq
\label{appDaea}
\tilde F_{tree}(\tilde t_+,t_-;t_1,t_2)=\lambda^{-1} \tilde t_+ t_s+
F_1^{(0)}(St_s+Ct_-,t_1)+F_2^{(0)}(Ct_s-St_-,t_2)
~,
\eeq
\beq
\label{appDaeb}
\tilde F_{one-loop}(\tilde t_+,t_-;t_1,t_2)=
F_1^{(1)}(St_s+Ct_-,t_1)+F_2^{(1)}(Ct_s-St_-,t_2)
-\frac{1}{2} \log \left[ S^2 \langle \PP_1 \PP_1 \rangle_0 +
C^2 \langle \PP_2 \PP_2 \rangle_0 \right]
~,
\eeq
\end{subequations}
where $F_i=\sum_{g=0}^\infty \lambda^{g-1} F_i^{(g)}$ $(i=1,2)$
is the genus expansion of the free energy of each theory.
We see that the tree-level expression \eqref{appDaea} is merely a Legendre
transform of the total undeformed free energy. On the other hand, the
one-loop expression \eqref{appDaeb} receives contributions both from
tree-level and one-loop quantities in the undeformed theory.

At any genus, correlation functions can be computed with the use of the Feynman
diagrams in fig.\ \ref{vertices}, which include vertices (tadpoles, 2- and higher
$n$-point vertices) of both minimal strings. Vertices of either string can be connected
with a common propagator. The strength of the vertices and the actual form
of the propagator is controlled by the double-trace parameter $\theta$.
The generic diagram of the modified theory is a sum of correlation functions
in the original product theory on disconnected worldsheets. One could reformulate
the new interactions in terms of a set of non-local interactions on the worldsheet
of the original theory. Because of its tractability, our setup provides a concrete
example of the non-local string theory construction proposed in \cite{nonlocal}.
We will return to this point in the next section.

To illustrate the generic structure of correlation functions in the modified product
theory with a few examples, we will now consider the two-point functions of the
operators $\PP_i$, $\OO_i$ $(i=1,2)$ at tree-level and one-loop.

\FIGURE{
\vspace{.4cm}
\centerline{\includegraphics[width=8.5cm,height=8.6cm]{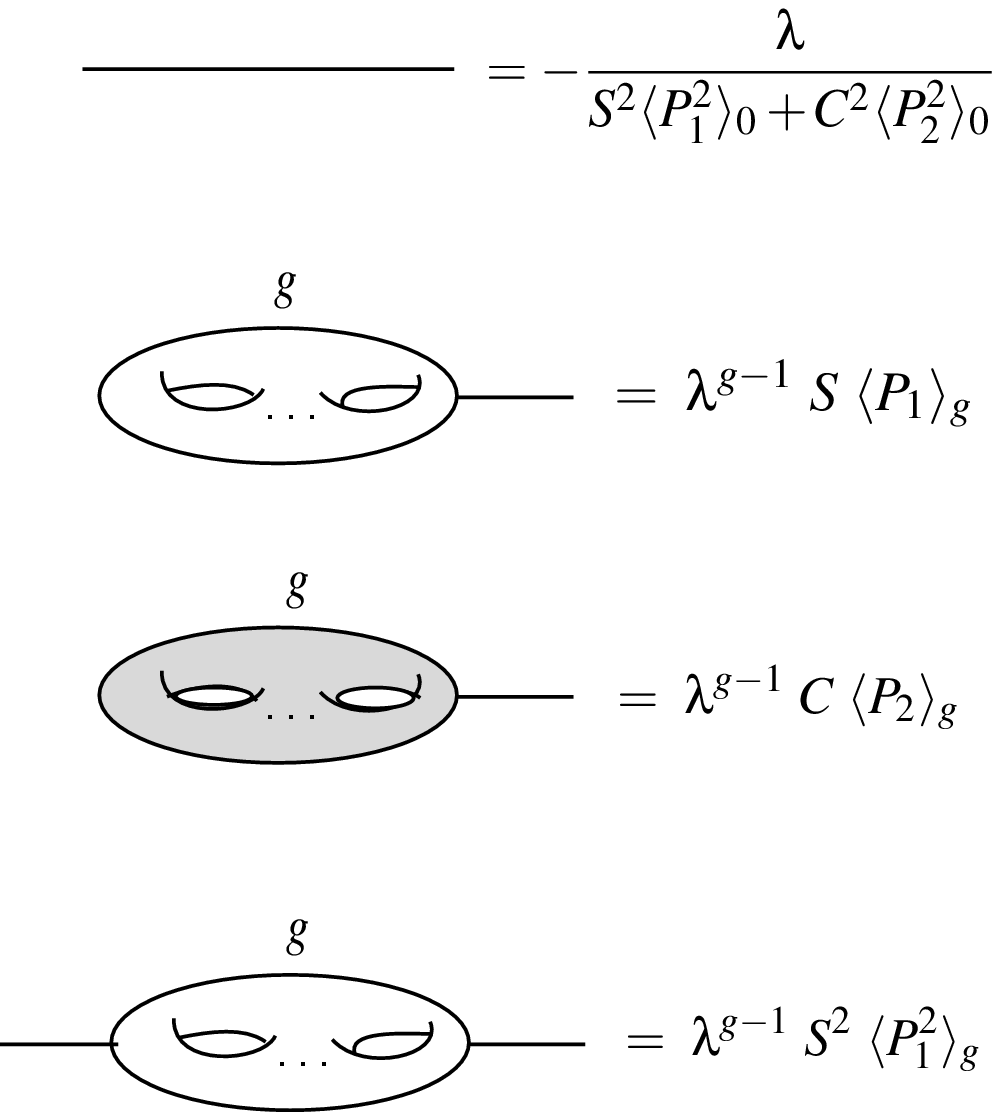}}
\vspace{.7cm}
\centerline{\includegraphics[width=8.3cm,height=1.6cm]{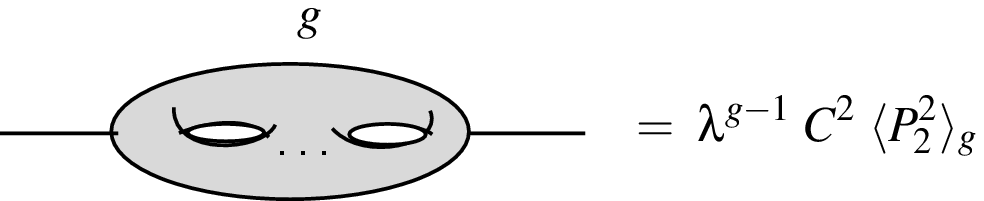}}
\vspace{.7cm}
\centerline{\includegraphics[width=14.5cm,height=3.7cm]{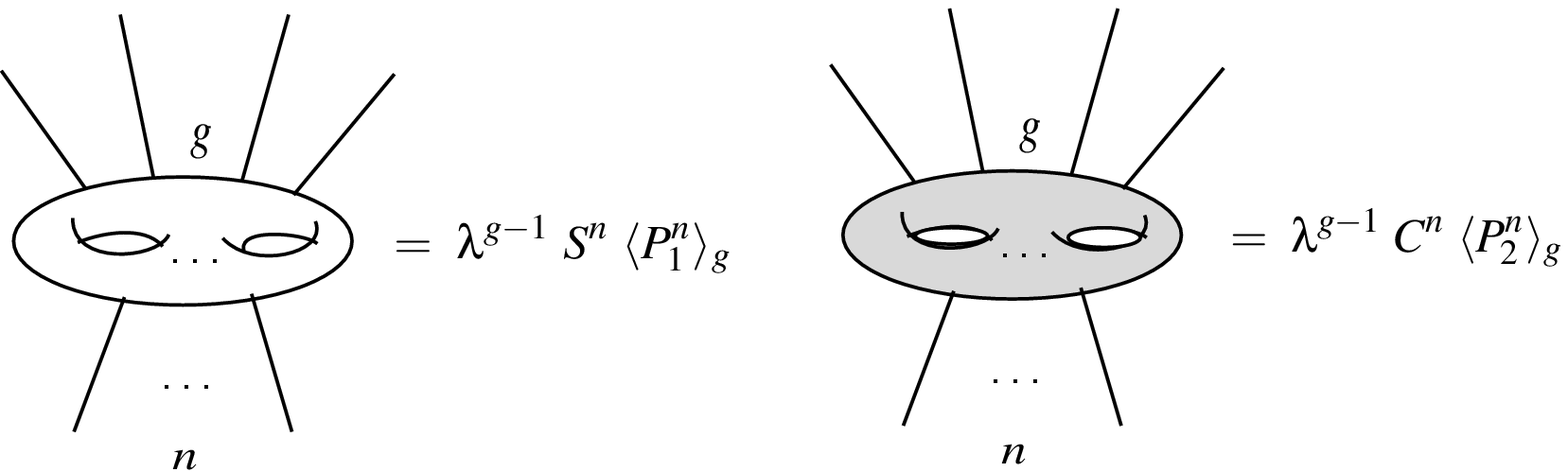}}
\caption{\small \it Vertices necessary for the diagrammatic expansion of the
modified free energy of two coupled minimal strings. The white (grey)
worldsheets are respectively worldsheets of minimal string 1 (2). The vertices
in the last line apply to any genus $g=0,1,2,...$ and can have $n=3,4,5,...$ external
legs.}
\label{vertices}
}

\subsection{Two-point functions at tree-level}

At tree-level the free energy of the modified theory is the Legendre transform
of the original theory. We can obtain the tree-level amplitudes of the scaling operators
$\OO_1$, $\OO_2$ by differentiating \eqref{appDaea} with respect to $t_1$, $t_2$
and using the chain rule. For two-point functions we find (we will denote the
modified 2-point functions as $\llangle \cdots \rrangle$)
\begin{subequations}
\beq
\label{appDba}
\llangle \OO_1 \OO_1 \rrangle_0= \langle \OO_1 \OO_1 \rangle_0+
\frac{S^2 \langle \OO_1 \PP_1\rangle_0^2}{S^2 \langle \PP_1 \PP_1 \rangle_0+
C^2 \langle \PP_2 \PP_2 \rangle_0}
~,
\eeq
\beq
\label{appDbb}
\llangle \OO_2 \OO_2 \rrangle_0= \langle \OO_2 \OO_2 \rangle_0+
\frac{C^2 \langle \OO_2 \PP_2\rangle_0^2}{S^2 \langle \PP_1 \PP_1 \rangle_0+
C^2 \langle \PP_2 \PP_2 \rangle_0}
~,
\eeq
\beq
\label{appDbc}
\llangle \OO_1 \OO_2 \rrangle_0=
\frac{SC \langle \OO_1 \PP_1\rangle_0 \langle \OO_2 \PP_2 \rangle_0}
{S^2 \langle \PP_1 \PP_1 \rangle_0+C^2 \langle \PP_2 \PP_2 \rangle_0}
~.
\eeq
\end{subequations}
As expected, the last correlation function that describes the non-trivial
interaction between the two minimal strings is proportional to $SC$, which vanishes
only when the double-trace coupling $g_{12}$ is zero.
Diagrammatically, these two-point functions can be recast as
\begin{subequations}
\beq
\label{appDbda}
\llangle \OO_1 \OO_1 \rrangle_0=~
\raisebox{-.4cm}{\includegraphics[width=4.3cm,height=.9cm]{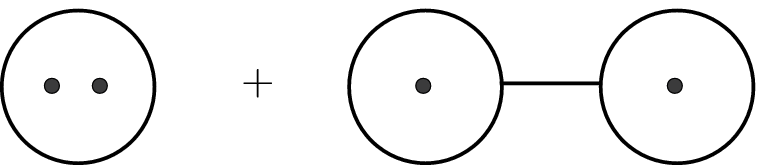}}
~,
\eeq
\beq
\label{appDbdb}
\llangle \OO_2 \OO_2 \rrangle_0=~
\raisebox{-.4cm}{\includegraphics[width=4.3cm,height=.9cm]{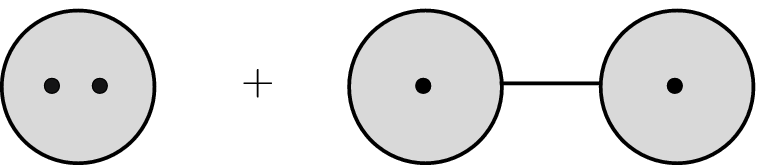}}
~,
\eeq
\beq
\label{appDbdc}
\llangle \OO_1 \OO_2 \rrangle_0=~
\raisebox{-.4cm}{\includegraphics[width=2.3cm,height=.9cm]{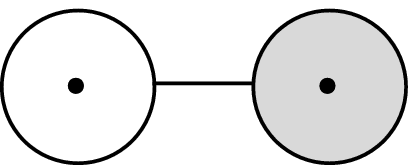}}
~.
\eeq
\end{subequations}

Differentiating the free energy $\tilde F$ with respect to the scaling parameters
$\tilde t_+$, $t_-$ gives correlation functions for the operators $\tilde \PP_+, \PP_-$,
which are defined in the following way. In the original theory, before the double-trace
deformation, $\PP_\pm$ are operators in the product theory defined by the linear
combinations
\beq
\label{appDbe}
\PP_+=S \PP_1+C\PP_2~, ~ ~ \PP_-=C \PP_1-S \PP_2
~.
\eeq
The couplings of $\PP_\pm$ are what we called above $t_\pm$.
In the deformed theory \eqref{appDaa}, $t_+$ is modified to $\tilde t_+$
and $t_-$ remains invariant. Accordingly, $\PP_+$ is modified to
$\tilde \PP_+$ and $\PP_-$ remains invariant.

At tree-level, $\tilde \PP_+$ and $\PP_-$ exhibit the one-point functions
\beq
\label{appDabf}
\llangle \tilde \PP_+ \rrangle_0=\lambda^{-1} t_s
~, ~~
\llangle \PP_- \rrangle_0=\lambda^{-1}\langle \PP_- \rangle_0
\eeq
and the two-point functions
\begin{subequations}
\beq
\label{appDabga}
\llangle \tilde \PP_+ \tilde \PP_+ \rrangle_0=
-\frac{1}{S^2 \langle \PP_1\PP_1\rangle_0 +C^2 \langle \PP_2 \PP_2 \rangle_0}
~,
\eeq
\beq
\label{appDabgb}
\llangle \PP_- \PP_- \rrangle_0=C^2 \langle \PP_1 \PP_1\rangle_0+
S^2\langle \PP_2 \PP_2 \rangle_0
-\frac{C^2 S^2 \left( \langle \PP_1 \PP_1 \rangle_0 - \langle \PP_2 \PP_2 \rangle_0\right)^2}
{S^2 \langle \PP_1\PP_1\rangle_0 +C^2 \langle \PP_2 \PP_2 \rangle_0}
~,
\eeq
\beq
\label{appDabgc}
\llangle \tilde \PP_+ \PP_- \rrangle_0=-
\frac{CS \left(\langle \PP_1 \PP_1 \rangle_0 - \langle \PP_2 \PP_2 \rangle_0\right)}
{S^2 \langle \PP_1\PP_1\rangle_0 +C^2 \langle \PP_2 \PP_2 \rangle_0}
~.
\eeq
\end{subequations}

\subsection{Two-point functions at one-loop}

One-loop correlation functions can be obtained from \eqref{appDaeb}
in a similar fashion. Here we will list the two-point functions of the
operators $\OO_i$. Analogous expressions can be deduced for
the amplitudes of the operators $\tilde \PP_+$, $\PP_-$. It will be convenient
to set
\beq
\label{kprop}
\KK=S^2 \langle \PP_1 \PP_1 \rangle_0+C^2\langle \PP_2 \PP_2 \rangle_0
~.
\eeq

The two-point function $\llangle \OO_1 \OO_1 \rrangle_1$ reads
\bea
\label{appDaca}
&&\llangle \OO_1 \OO_1 \rrangle_1=\langle \OO_1 \OO_1 \rangle_1+
S^2 \KK^{-1} \langle \PP_1 \OO_1\rangle_0 \langle \PP_1 \OO_1 \rangle_1
-S^6 \KK^{-3} \langle \PP_1 \rangle_1 \langle \PP_1 \OO_1 \rangle_0^2
\langle\PP_1^3\rangle_0-
\nonumber\\
&&- C^3 S^3 \KK^{-3} \langle \PP_1 \rangle_1 \langle \PP_1 \OO_1 \rangle_0^2
\langle\PP_2^3\rangle_0
-CS^5 \KK^{-3} \langle \PP_2 \rangle_1 \langle \PP_1 \OO_1 \rangle_0^2
\langle\PP_1^3\rangle_0
-C^3S^2 \KK^{-3} \langle \PP_2 \rangle_1 \langle \PP_1 \OO_1 \rangle_0^2
\langle\PP_2^3\rangle_0
+\nonumber\\
&&+ S^4 \KK^{-2}\langle \PP_1^2 \rangle_1 \langle \PP_1 \OO_1 \rangle_0^2
+CS^3 \KK^{-2}\langle \PP_2^2 \rangle_1 \langle \PP_1 \OO_1 \rangle_0^2
+S^2 \KK^{-1} \langle \PP_1 \rangle_1 \langle \PP_1 \OO_1^2 \rangle_0
+CS \KK^{-1} \langle \PP_2 \rangle_1 \langle \PP_1 \OO_1^2 \rangle_0
+\nonumber\\
&&+\frac{1}{2}S^4  \KK^{-2} \langle \PP_1^2 \OO_1 \rangle_0
+ S^6 \KK^{-3} \langle \PP_1^2 \OO_1 \rangle_0 \langle \PP_1 \OO_1 \rangle_0
\langle \PP_1^3 \rangle_0
+ C^3S^3 \KK^{-3} \langle \PP_1^2 \OO_1 \rangle_0 \langle \PP_1 \OO_1 \rangle_0
\langle \PP_2^3 \rangle_0
-\nonumber\\
&&-\frac{1}{2} S^2 \KK^{-1} \langle \PP_1^2 \OO_1^2 \rangle_0
-S^4 \KK^{-2} \langle \PP_1 \OO_1 \rangle_0 \langle \PP_1^3 \OO_1 \rangle_0
-\frac{1}{2} S^4 \KK^{-2} \langle \PP_1^3 \rangle_0 \langle \PP_1 \OO_1^2 \rangle_0
-\nonumber\\
&&-\frac{1}{2} CS^3 \KK^{-2} \langle \PP_2^3 \rangle_0 \langle \PP_1 \OO_1^2 \rangle_0
-\frac{1}{2} S^6 \KK^{-3} \langle \PP_1 \OO_1 \rangle_0^2 \langle \PP_1^4 \rangle_0
-\frac{1}{2} C^4 S^2 \KK^{-3} \langle \PP_1 \OO_1 \rangle_0^2 \langle \PP_2^4 \rangle_0
+\nonumber\\
&&+S^8 \KK^{-4} \langle \PP_1 \OO_1 \rangle_0^2 \langle \PP_1^3 \rangle_0^2
+ 2 C^3S^5 \KK^{-4} \langle \PP_1 \OO_1 \rangle_0^2 \langle \PP_1^3 \rangle_0
\langle \PP_2^3 \rangle_0
+C^6 S^2 \KK^{-4} \langle \PP_1 \OO_1 \rangle_0^2 \langle \PP_2^3 \rangle_0^2
\eea
and has the diagrammatic expansion in fig.\  \ref{o1o1feyn}.
A similar expansion, with obvious modifications, applies to
$\llangle \OO_2 \OO_2 \rrangle_1$.

\FIGURE{
\vspace{.4cm}
\centerline{\includegraphics[width=14cm,height=.8cm]{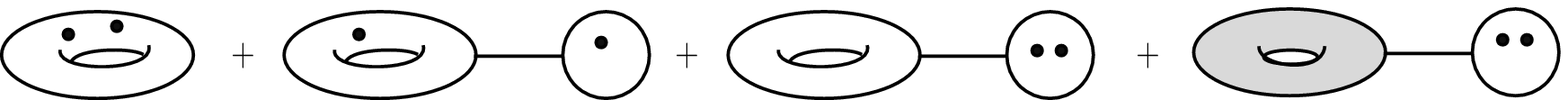}}
\vspace{.7cm}
\centerline{\includegraphics[width=11.8cm,height=.8cm]{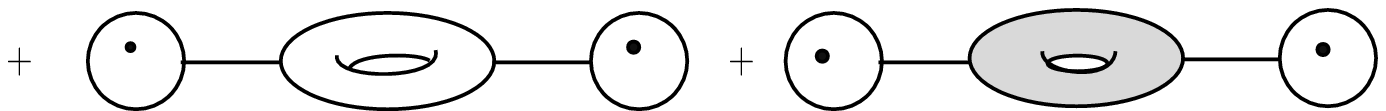}}
\vspace{.7cm}
\centerline{\includegraphics[width=14cm,height=.9cm]{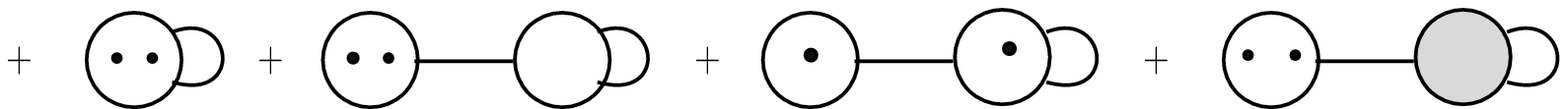}}
\vspace{.7cm}
\centerline{\includegraphics[width=14cm,height=.8cm]{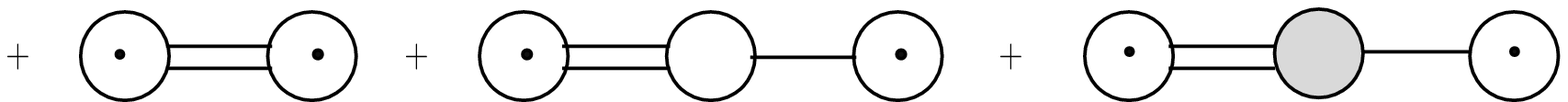}}
\vspace{.7cm}
\centerline{\includegraphics[width=14cm,height=1.9cm]{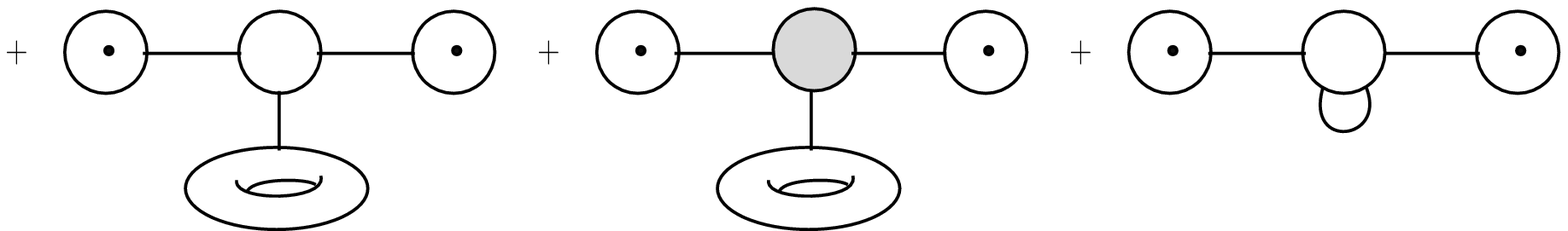}}
\vspace{.7cm}
\centerline{\includegraphics[width=14cm,height=1.9cm]{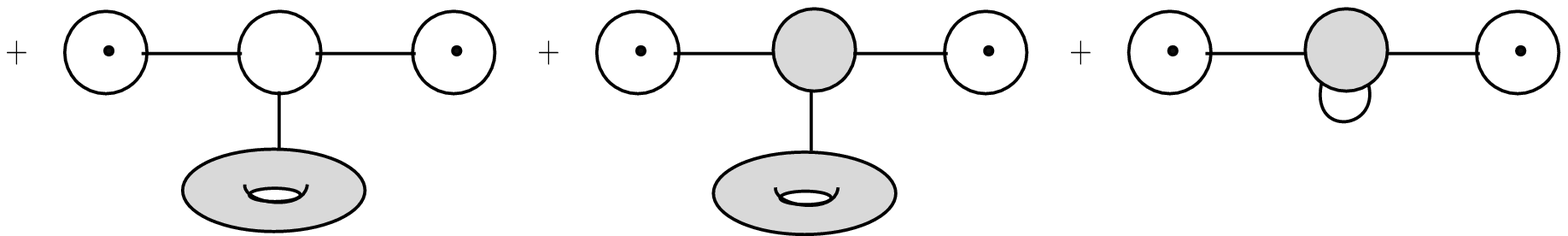}}
\vspace{.7cm}
\centerline{\includegraphics[width=14cm,height=2.1cm]{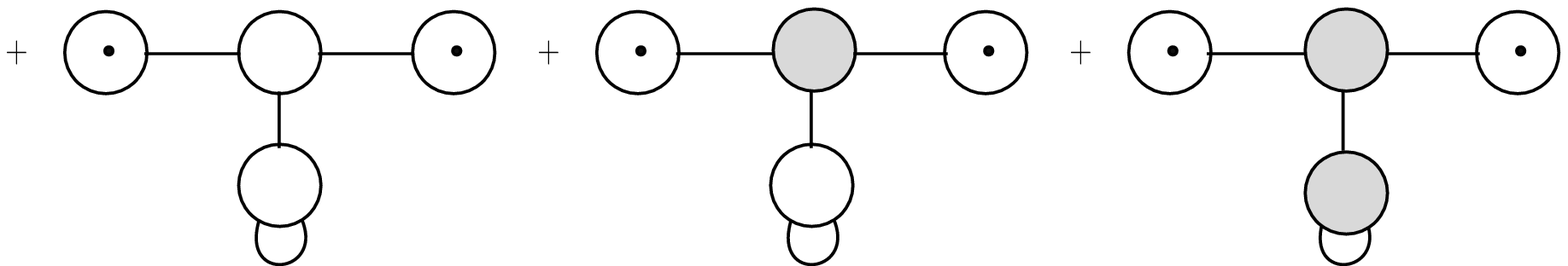}}
\caption{\small \it The diagrammatic expansion of the one-loop amplitude
$\llangle \OO_1 \OO_1 \rrangle_1$.}
\label{o1o1feyn}
}

The expansion of $\llangle \OO_1 \OO_2 \rrangle_1$ is
\bea
&&\llangle \OO_1 \OO_2 \rrangle_1=CS \KK^{-1}
\langle \PP_1 \OO_1 \rangle_1\langle \PP_2 \OO_2 \rangle_0
+CS \KK^{-1} \langle \PP_1 \OO_1 \rangle_0 \langle \PP_2 \OO_2 \rangle_1
-\nonumber\\
&&-CS^5 \KK^{-3} \langle \PP_1 \rangle_1 \langle \PP_1 \OO_1 \rangle_0
\langle \PP_2 \OO_2 \rangle_0  \langle \PP_1^3 \rangle_0
-C^2S^4 \KK^{-3} \langle \PP_2 \rangle_1 \langle \PP_1 \OO_1 \rangle_0
\langle \PP_2 \OO_2 \rangle_0  \langle \PP_1^3 \rangle_0
-\nonumber\\
&&-C^4S^2 \KK^{-3} \langle \PP_1 \rangle_1 \langle \PP_1 \OO_1 \rangle_0
\langle \PP_2 \OO_2 \rangle_0  \langle \PP_2^3 \rangle_0
-C^5S \KK^{-3} \langle \PP_2 \rangle_1 \langle \PP_1 \OO_1 \rangle_0
\langle \PP_2 \OO_2 \rangle_0  \langle \PP_2^3 \rangle_0
-\nonumber\\
&&-C^2 S^2 \KK^{-2} \langle \PP_1 \rangle_1 \langle \PP_1 \OO_1 \rangle_0
\langle \PP_2^2 \OO_2 \rangle_0
-C^3 S \KK^{-2} \langle \PP_2 \rangle_1 \langle \PP_1 \OO_1 \rangle_0
\langle \PP_2^2 \OO_2 \rangle_0
+\nonumber\\
&&+CS^3 \KK^{-2} \langle\PP_1 \OO_1 \rangle_0 \langle \PP_2 \OO_2 \rangle_0
\langle \PP_1^2 \rangle_1
+C^3S \KK^{-2} \langle\PP_1 \OO_1 \rangle_0 \langle \PP_2 \OO_2 \rangle_0
\langle \PP_2^2 \rangle_1
+\nonumber\\
&&+CS^3 \KK^{-2} \langle \PP_1 \rangle_1 \langle \PP_1^2 \OO_1 \rangle_0
\langle \PP_2 \OO_2 \rangle_0
+C^2 S^2 \KK^{-2} \langle \PP_2 \rangle_1 \langle \PP_1^2 \OO_1 \rangle_0
\langle \PP_2 \OO_2 \rangle_0
+\nonumber\\
&&+\frac{1}{2}C^2 S^2 \KK^{-2}  \langle \PP_1^2 \OO_1\rangle_0
\langle \PP_2^2 \OO_2 \rangle_0
-\frac{1}{2} C S^3 \KK^{-2} \langle \PP_1^3 \OO_1 \rangle_0
\langle \PP_2 \OO_2 \rangle_0
-\frac{1}{2} C^3 S \KK^{-2} \langle \PP_1 \OO_1 \rangle_0
\langle \PP_2^3 \OO_2 \rangle_0
+\nonumber
\eea
\bea
\label{appDacbi}
&&+C S^7 \KK^{-4} \langle \PP_1 \OO_1\rangle_0
\langle \PP_2\OO_2 \rangle_0 \langle \PP_1^3 \rangle_0
+ C^7 S \KK^{-4} \langle \PP_1 \OO_1\rangle_0
\langle \PP_2\OO_2 \rangle_0 \langle \PP_2^3 \rangle_0
+\nonumber\\
&&+2 C^4 S^4 \KK^{-4} \langle \PP_1 \OO_1\rangle_0
\langle \PP_2\OO_2 \rangle_0 \langle \PP_1^3 \rangle_0 \langle \PP_2^3 \rangle_0
+ C^2 S^4 \KK^{-3} \langle \PP_1 \OO_1 \rangle_0
\langle \PP_2^2 \OO_2 \rangle_0 \langle \PP_1^3 \rangle_0
-\nonumber\\
&&-\frac{1}{2} C S^5 \KK^{-3} \langle \PP_1 \OO_1 \rangle_0
\langle \PP_2 \OO_2 \rangle_0 \langle \PP_1^4 \rangle_0
-\frac{1}{2} C^5 S \KK^{-3} \langle \PP_1 \OO_1 \rangle_0
\langle \PP_2 \OO_2 \rangle_0 \langle \PP_2^4 \rangle_0
~.
\eea
The corresponding diagrams appear in fig.\ \ref{o1o2feyn}.

Most of the terms appearing in equations \eqref{appDaca}
and \eqref{appDacbi} have a unique representation as diagrams in
figures \ref{o1o1feyn} and \ref{o1o2feyn}. There are, however,
some exceptions. These include the last three diagrams in fig.\
\ref{o1o1feyn} and their analogs in fig.\ \ref{o1o2feyn}. For example,
both diagrams in fig.\ \ref{o1o1alter} can be associated to the
amplitude contribution
$\langle \PP_1 \OO_1\rangle_0^2 \langle \PP_1^3\rangle_0^2$.
Viewing the contact interactions as tiny wormholes that connect
different parts of a worldsheet, we recognize these diagrams as
representations of the same amplitude in a different region of
the worldsheet moduli. In string perturbation theory one sums over
all these configurations automatically.

There are many contributions to the one-loop renormalization of
the two-point functions \eqref{appDbda} -- \eqref{appDbdc}. Some
of them are standard torus amplitudes in theory 1 or 2 separately,
but the majority comes from contact interactions where worldsheets
of theory 1 and/or 2 connect via a tiny `neck' (wormhole) associated
to a common propagator $\KK^{-1}$ \eqref{kprop}. Contact interactions
between worldsheets of different theories are especially interesting
since they provide a clear picture of how theories 1 and 2
communicate with each other on the level of the worldsheet.
Such interactions contribute both to single theory two-point functions
($\llangle \OO_1 \OO_1 \rrangle$ or $\llangle \OO_2 \OO_2 \rrangle$)
and to mixed two-point functions of the form $\llangle \OO_1 \OO_2 \rrangle$.
In the latter case the 1-2 interactions work to renormalize the tree-level
result \eqref{appDbdc} either by renormalizing separately the single
theory two-point functions $\langle \OO_1 \PP_1 \rangle_0$,
$\langle \OO_2 \PP_2 \rangle_0$ or by renormalizing the propagator
$\KK^{-1}$ with a double wormhole interaction
$C^2S^2  \KK^{-2} \langle \PP_1^2 \OO_1\rangle_0 \langle \PP_2^2\OO_2 \rangle$
(the last diagram in the third line of fig.\ \ref{o1o2feyn}). A similar diagram
renormalizes the mixed two-point function $\langle g_1 g_2\rangle$ for
the two AdS gravitons $g_1,g_2$ in the AdS/CFT
example of \cite{kir,ack}. In that case, this is precisely the effect that lifts the mass
of a certain linear combination of the gravitons and leads to massive gravity.

\FIGURE{
\vspace{.4cm}
\centerline{\includegraphics[width=8.5cm,height=.9cm]{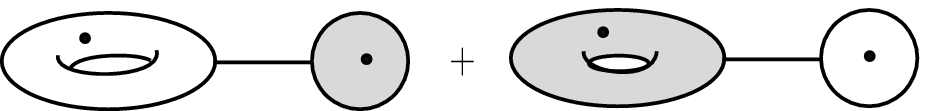}}
\vspace{.7cm}
\centerline{\includegraphics[width=12cm,height=.85cm]{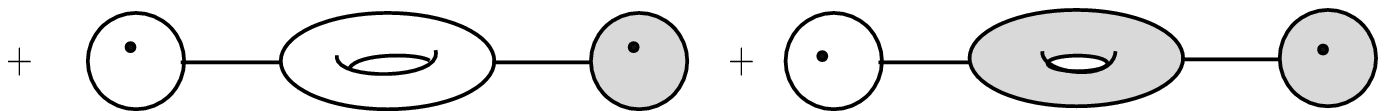}}
\vspace{.7cm}
\centerline{\includegraphics[width=11.5cm,height=.85cm]{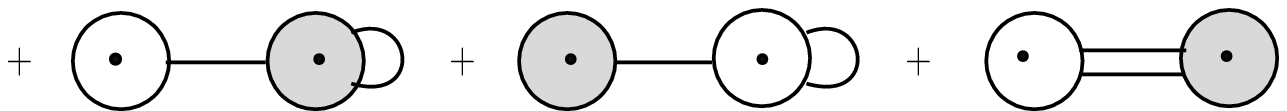}}
\vspace{.7cm}
\centerline{\includegraphics[width=12cm,height=1.9cm]{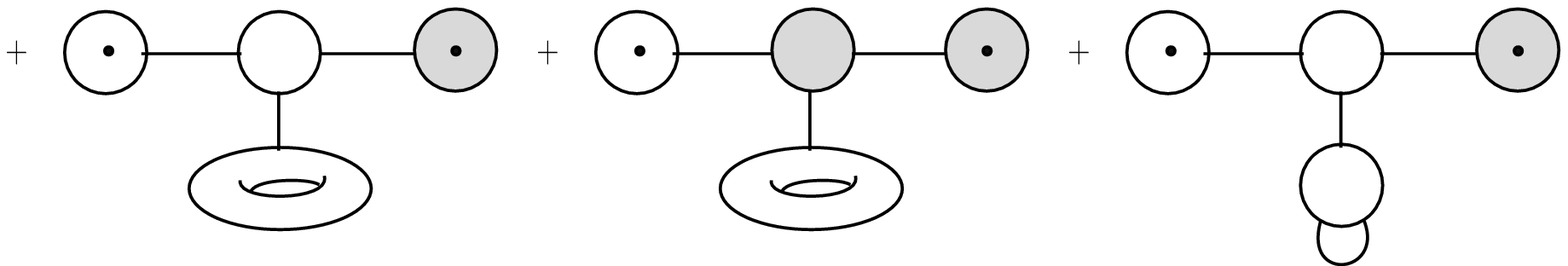}}
\vspace{.7cm}
\centerline{\includegraphics[width=12cm,height=1.9cm]{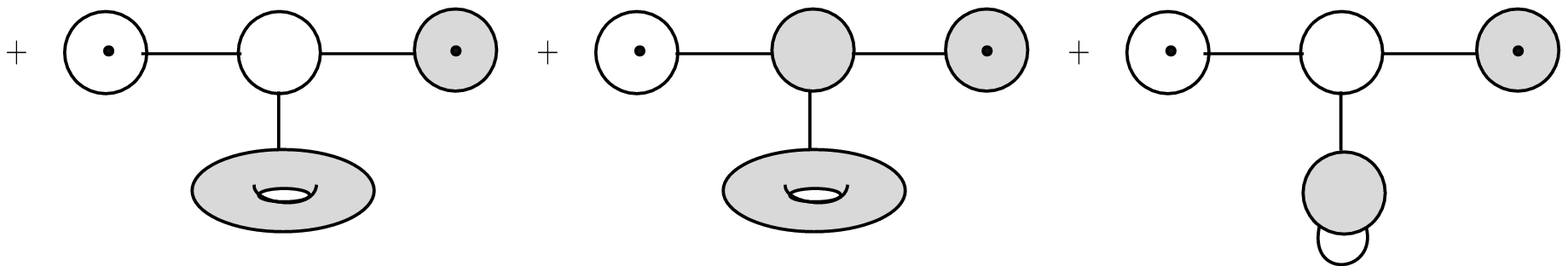}}
\vspace{.7cm}
\centerline{\includegraphics[width=12cm,height=1.9cm]{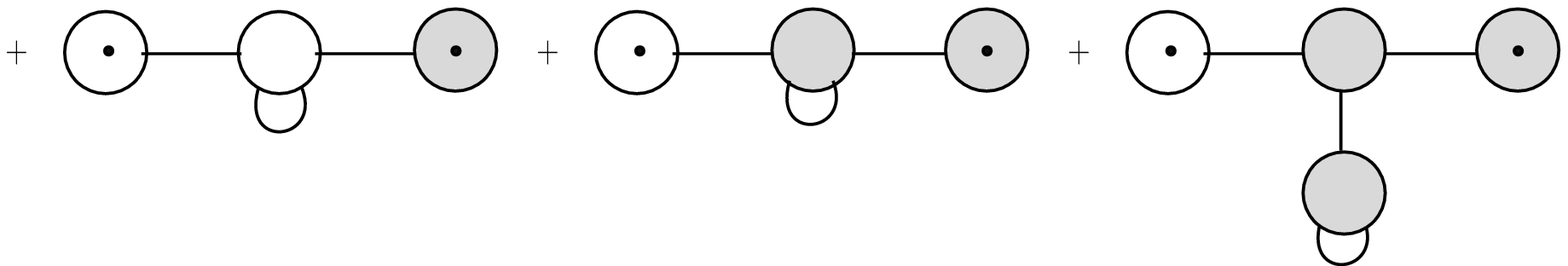}}
\vspace{.7cm}
\centerline{\includegraphics[width=12cm,height=0.6cm]{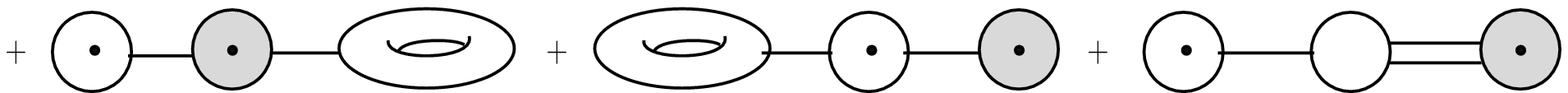}}
\vspace{.7cm}
\centerline{\includegraphics[width=12cm,height=0.7cm]{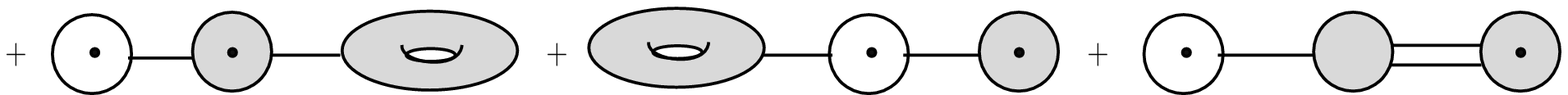}}
\caption{\small \it The diagrammatic expansion of the one-loop amplitude
$\llangle \OO_1 \OO_2 \rrangle_1$.}
\label{o1o2feyn}
}

\

\FIGURE{
\centerline{
\includegraphics[width=3.1cm,height=1.7cm]{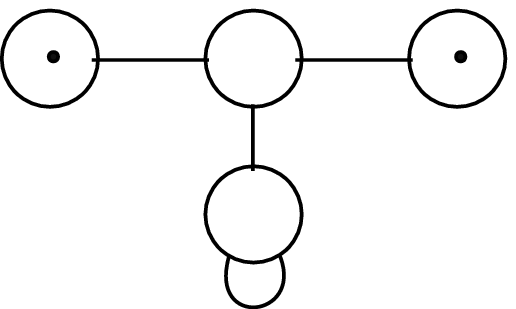}
~~~~~~~~~
\raisebox{.6cm}{\includegraphics[width=4.2cm,height=.65cm]{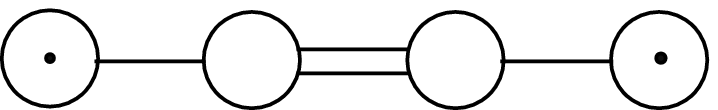} }
}
\caption{\small \it Two diagrams corresponding to the same amplitude contribution
$\langle \PP_1 \OO_1\rangle_0^2 \langle \PP_1^3\rangle_0^2$.}
\label{o1o1alter}
}

\section{Worldsheet and spacetime aspects of the multi-trace coupling}
\label{spacetime}

In this section we want to discuss the worldsheet and spacetime
interpretation of the theories defined above through the matrix
model. There are two aspects we want to emphasize: $(a)$ the
modified genus expansion and non-locality on the worldsheet, and
$(b)$ the analogy with higher dimensional AdS/CFT examples and
the possibility of an alternative formulation of the deformed theories
that does not involve non-locality on the worldsheet.

\subsection{Touching surfaces in Liouville gravity and the NLST structure}

In conventional matrix models, $i.e.$ matrix models with only single-trace
interactions, the Feynman diagram expansion generates discretized random
surfaces from which the continuous string worldsheet arises in a suitable
double scaling limit. Multi-trace interactions modify this expansion by gluing
together different random surfaces at a plaquette. In the continuum limit,
one can think of this contact interaction as a tiny neck (wormhole) that creates
a network of touching surfaces. Such microscopic degenerations of the string
worldsheet are already present in the conventional theory without the multi-trace
deformation, however, they give a sizable contribution that modifies the
conventional result only for special values of the multi-trace couplings where
a new double scaling limit is possible. The diagrammatic expansion of
the previous section gives a clear demonstration of this effect of contact
interactions between different worldsheets. In fact, the examples of section
\ref{correlations} give an even more dramatic illustration of this general
phenomenon: even worldsheets of different string theories can touch via
a common wormhole.

One can argue on general grounds that multi-trace interactions
will have a similar effect also in the AdS/CFT correspondence in higher
dimensions. String theories resulting from such a deformation were termed
non-local string theories in \cite{nonlocal}, since one can presumably
reproduce the effect of the tiny wormholes with a non-local deformation of
the worldsheet action. More specifically, if $N_w$ is the number of disconnected
components of the worldsheet $\Sigma=\Sigma^{(1)} \oplus \Sigma^{(2)} \oplus \cdots
\oplus \Sigma^{(N_w)}$ one can envision a non-local worldsheet action of the
form
\bea
\label{spaceaa}
&&\SS_{ws}=\sum_{i=1}^{N_w} \int d^2 \sigma^{(i)} \sqrt{g(\sigma)} \LL_0+
\sum_{i,j=1}^{N_w}\int d^2 \sigma_1^{(i)}d^2 \sigma_2^{(j)}
\sqrt{g((\sigma_1)g(\sigma_2)} \GG[X(\sigma_1),X(\sigma_2)]+
\nonumber\\
&&+trilocal~ and ~ higher-order~ interactions
~~
\eea
where $X$ denotes collectively the target space fields and $\GG$ is a non-local
interaction. $\GG$ and the rest of the higher order interactions in \eqref{spaceaa}
should be determined directly in string theory by consistency, $e.g.$ by requiring
the cancellation of the Weyl anomaly. The precise rules for such theories have not
been worked out, however, the consistency of the AdS/CFT correspondence implies
that such non-local string theories should exist. In fact, we recognize a concrete
example of such a theory in the non-critical NLSTs defined in section
\ref{sec:matrixmodels}. In that case, see section \ref{correlations}, we can
determine explicitly the precise rules for the diagrammatic expansion of the
correlation functions and from them one could work backwards to deduce the
required form of $\GG$ $etc$.

\subsection{An alternative interpretation?}

When a CFT admits a dual supergravity description, it has been argued
\cite{witten,berkooz,muck,minces,petkou} that there is an alternative
way to understand multi-trace deformations as mixed boundary conditions
for the dual fields in AdS. In the matrix model case, there are indications
for a similar reformulation of the deformation on the string theory side as
a local effect on the worldsheet that involves changing the
branch of the Liouville dressing of a vertex operator \cite{Klebanov:1994pv}.
Here we will argue that there is a simple interpretation of this observation
at tree-level, which is in accordance with our current understanding of Liouville
theory and the analogous observations in the AdS/CFT correspondence
\cite{witten,berkooz,muck,minces,petkou}. Beyond tree-level, a similar
reformulation does not go through, and one has to think of the deformed
theory as an NLST along the lines of \cite{nonlocal}.

\subsubsection{The tree-level theory}
\label{treelevel}

To illustrate the main point, let us consider first the simpler situation
of a single one-matrix model deformed by a double-trace deformation.
The analogous AdS/CFT example involves a CFT deformed by a
double-trace interaction of the form
\beq
\label{spaceba}
\delta \SS=\int d^dx ~ g \OO^2
~,
\eeq
where $\OO$ is a single-trace operator with scaling dimension $\Delta$.
$g$ is a coupling that scales like $N^0$. It will be useful to recall first the
main features of this example.

If the scaling dimension $\Delta$ is less than $\frac{d}{2}$, the perturbing
operator \eqref{spaceba} is relevant and the theory runs towards an infrared (IR)
fixed point, where $\OO$ has a different scaling dimension $d-\Delta$. At large
$N$, the IR theory is simply related to the UV theory by Legendre transform
\cite{Gubser:2002vv}. At tree-level and within the supergravity description,
this RG running on the boundary has a holographic counterpart in
$(d+1)$-dimensional AdS space as a flow between different boundary
conditions for the bulk field $\phi$ which is dual to the operator $\OO$.

There are two points we want to
emphasize with respect to this statement. First, in contrast with the case
of a single-trace perturbation, where the dual gravity background is
deformed away from AdS, the multi-trace deformation does not backreact
to the bulk geometry to leading order in  $1/N$. Second, changing the boundary
conditions of the bulk field does not mean that we pick a different solution
of the bulk equations of motion for the dual field $\phi$. This is still controlled
by regularity in the bulk. Instead, we modify the definition of
the source -- in other words, we modify the bulk/boundary dictionary.
This last point will be important for the matrix model discussion below,
so we take a moment to explain it here in slightly more detail (for a more
detailed exposition of this well-known material we refer the reader to
\cite{Hartman:2006dy}, which is what we will follow here mostly, and the
standard references therein).

In coordinates where the Euclidean AdS metric takes the form
\beq
\label{spacebb}
ds^2_{d+1}=\frac{1}{r^2}\left(dr^2+\sum_{i=1}^{d}(dx^i)^2\right)
\eeq
consider a scalar field $\phi$ with the action
\beq
\label{spacebc}
S=\int d^d x dr \sqrt g \left(\frac{1}{2}(\d \phi)^2+\frac{1}{2}m^2 \phi^2\right)+
\int d^d x \sqrt g ~ \LL_{boundary}|_{r=\epsilon}
~.
\eeq
$\phi$ is the scalar field dual to the single-trace operator $\OO$. The mass
$m$ is related to the scaling dimension $\Delta$ via the relation
\beq
\label{spacebd}
\Delta=\frac{d}{2}-\nu~, ~ ~ \nu \equiv \sqrt{\frac{d^2}{4}+m^2}
~.
\eeq
We have included a boundary term to the action \eqref{spacebc}, which
is defined at $r=\epsilon$, the IR bulk regulator. To reproduce the double-trace
deformation \eqref{spaceba} we are instructed to use the boundary
interaction
\beq
\label{spacebe}
\LL_{boundary}=\frac{1}{2} f \phi^2
~.
\eeq
$f$ has a simple relation with the field theory $g$ in \eqref{spaceba}.
A careful treatment gives
\beq
\label{spacebf}
f=-\Delta-2g\epsilon^{2\nu}\left(2\pi^{d/2}
\frac{\Gamma(1-\nu)}{\Gamma(\Delta)}\right)
~.
\eeq
Varying the action \eqref{spacebc} with respect to $\phi$ we deduce the
mixed Neumann/Dirichlet boundary conditions
\beq
\label{spacebg}
f~ \phi(x,\epsilon)+\d \phi(x,\epsilon)\cdot \hat n=0
~,
\eeq
where $\hat n=\epsilon \hat r$ is a unit vector specifying the normal to the
boundary.

To compute correlation functions in the AdS/CFT correspondence,
we evaluate the on-shell bulk action as a functional of the boundary source
$\phi_b$. With mixed boundary conditions \eqref{spacebg}, the boundary
value problem is
\begin{subequations}
\bea
\label{spacebia}
(\Box-m^2)\phi(x,r)&=&0
~,\\
\label{spacebib}
f~ \phi(x,\epsilon)+\d \phi(x,\epsilon)\cdot \hat n&=&\phi_b(x)
~.
\eea
\end{subequations}
After the Fourier transformation
\beq
\label{spacebj}
\phi(x_i,r)=\frac{1}{(2\pi)^{d/2}}\int d^dk ~ e^{ik_i x_i} \phi(k_i,r)
\eeq
the wave equation \eqref{spacebia} becomes (set $\phi(k,r)\equiv r^{d/2} \chi(k,r)$)
\beq
\label{spacebk}
\left \{ -\left( r \frac{d}{dr}\right)^2+k^2 r^2+\nu^2 \right\}
\chi(k,r)=0
~.
\eeq
The unique solution to \eqref{spacebk}, \eqref{spacebib}, which is regular
at $r\to \infty$, is given by the modified Bessel function of the 2nd kind $\KK_\nu$.
Consequently,
\beq
\label{spacebl}
\phi(k,r)=\left( \frac{\phi_b(k)}{f~ \psi(k,\epsilon)+\d \psi(k,\epsilon)\cdot \hat n}
\right) \psi(k,r)~, ~ ~
\psi(k,r)\equiv r^{d/2} \KK_\nu(kr)
~.
\eeq
In particular, we notice that for any bare value of the double-trace coupling $g$,
the solution is always the same Bessel function. What changes is the relation
between the source $\phi_b$ and the asymptotic coefficients $\alpha,\beta$ in the
near boundary ($r\to 0$) expansion of $\phi$
\beq
\label{spacebm}
\phi(x,r) \sim r^{d-\Delta}\left[ \alpha(x)+\OO(r^2)\right]+
r^\Delta\left[ \beta(x)+\OO(r^2)\right]
~.
\eeq
Regularity, or equivalently the fact that we chose $\KK_\nu$ as the solution of
\eqref{spacebk}, has already fixed a linear relation between $\alpha$ and
$\beta$.

Having said this, let us return to the matrix model case and the dual minimal
strings. The analog of the wave equation \eqref{spacebk} is the Wheeler-DeWitt
(WdW) equation. For a minimal string with Liouville interaction
\beq
\label{spacebn}
\delta \SS_{Liouville}=\mu \int d^2z~ \OO_{matter}e^{\alpha_+ \varphi}
~, ~ ~ \alpha_+=b>0
\eeq
the wavefunction $\Psi_\nu$ for a mode that corresponds to the vertex operator
($Q$ is the linear dilaton slope)
\beq
\label{spacebo}
\VV_{\nu}=\OO_{matter} e^{-\frac{1}{2}(b\nu+Q) \varphi}
=\Psi_\nu e^{-\frac{Q}{2}\varphi}
\eeq
obeys, in the mini-superspace approximation, the WdW equation
\beq
\label{spacebp}
\left\{ -\left( \ell \frac{d}{d\ell}\right)^2+4\mu \ell^2+\nu^2 \right\} \Psi_\nu(\ell)
=0
~.
\eeq
$\ell$ is related to the zero-mode of the Liouville coordinate $\varphi_0$
by the relation
\beq
\label{spacebq}
\ell=e^{\frac{1}{2}b \varphi_0}
~.
\eeq
The weak coupling end of Liouville theory lies at $\ell \to 0$, or equivalently
$\varphi \to -\infty$.

The similarity between eqs.\ \eqref{spacebk} and \eqref{spacebp} is obvious.
Again, the only solution that is regular in the ``IR'', $i.e.$ the strong coupling
region at $\ell \to \infty$, is
\beq
\label{spacebr}
\Psi_\nu(\ell)\propto \KK_\nu(2\sqrt \mu \ell)
~.
\eeq
Writing $\KK_\nu$ in terms of the modified Bessel functions of the 1st kind
$\II_\nu$
\beq
\label{spacebs}
\KK_\nu=\frac{\pi}{2\sin(\pi \nu)} \left(\II_{-\nu}-\II_\nu\right)
\eeq
we observe the weak coupling asymptotics
\beq
\label{spacebt}
\KK_\nu(\ell) \sim \frac{\pi}{2\sin(\pi \nu)}\left( \frac{2^\nu}{\Gamma(1-\nu)} \ell^\nu
+ ...-\frac{2^{-\nu}}{\Gamma(1+\nu)} \ell^{-\nu} +...\right)
~.
\eeq
Hence, regularity of the full wavefunction requires the presence of both the `right'
branch with $\nu<0$, that satisfies the Seiberg bound \cite{Seiberg:1990eb},
and the `wrong' branch with $\nu>0$, with a fixed reflection relation.

Deforming the matrix model with a double-trace operator, $e.g.$
the operator $(\tr\Phi^4)^2$, one finds two possible double-scaling limits:
the standard one at double scaling parameter $g=0$, and another one with
different string susceptibility exponents at a special non-vanishing value of $g$
\cite{Klebanov:1994pv,Klebanov:1994kv}. These two critical points are analogous
to the UV and IR fixed points of the CFT deformation in \eqref{spaceba}.
The analysis of the string susceptibility exponents \cite{Klebanov:1994pv}
and tree-level correlation functions \cite{Barbon:1995dx} in the new
double scaling limit indicates that one can reproduce the matrix model
results in a Goulian-Li approach \cite{Goulian:1990qr} to the minimal
string by changing the Liouville dressing, from the right branch to the wrong
branch, of the vertex operator that is dual to the matrix model operator involved
in the double-trace deformation. Ref.\ \cite{Klebanov:1994pv} proposed
this as a possible interpretation of the string theory dual to the deformed
matrix model.

The statement about changing the branch of the Liouville dressing
cannot mean that we modify the boundary conditions of the WdW
equation in such a way that the new solution is solely $\II_{|\nu|}$.
This solution is singular in the strong coupling region $(\ell \to \infty)$,
and one would need additional ingredients to explain this singularity, $e.g.$
the singularity could be accounted for by the presence of a large number of
ZZ branes \cite{Zamolodchikov:2001ah} localized at the strong coupling region
(see \cite{Kutasov:2004fg} for a related discussion).
It is unclear, however, why ZZ branes would all of a sudden appear in the
minimal string as we change $g$ continuously from the undeformed point
to the new critical point. Also, this interpretation would not fit with our
understanding of the analogous situations in the AdS/CFT correspondence,
as described above.

Along the same lines, in recent years it has been understood that in Liouville
theory both the standard Liouville interaction \eqref{spacebn} and its dual
\beq
\label{spacebu}
\widetilde{\delta \SS}_{Liouville}=\tilde \mu \int d^2 z~
\OO_{matter}e^{\alpha_- \varphi}~, ~~
\alpha_-=-\left(Q+\frac{b}{2}\right)
\eeq
have to be present with a fixed relation $\tilde \mu=\tilde \mu(\mu)$.
Otherwise one cannot explain the structure of the exact two- and three-point
functions \cite{Dorn:1992at,Dorn:1994xn,Zamolodchikov:1995aa,
Teschner:2001rv,Teschner:2003en}.

In view of these observations, it is more appropriate to think of the deformed
theory at tree-level in the following way. As a specific example, let us consider the
case of the 2nd multi-critical matrix model deformed by $g(\tr\Phi^4)^2$ at the new critical
point where the partition function is given by \eqref{mmsingle}. At tree-level,
we can still think of the string theory dual of this matrix model as the usual (2,3)
minimal string -- the same as that at the undeformed $g=0$ point. We should,
however, modify the bulk/boundary dictionary and think of the amplitudes of
the new theory as a function of the $dual$ cosmological constant $\tilde \mu$.
A way to rephrase this proposal is as follows. It is known that Liouville theory
is symmetric under the simultaneous exchange of $\mu \leftrightarrow \tilde \mu$
and $\alpha_+ \leftrightarrow \alpha_-$ in \eqref{spacebn}, \eqref{spacebu}.
Performing half of this transformation, $e.g.$ $\mu \leftrightarrow \tilde \mu$
with $\alpha_+$ unchanged, gives a different theory. This is the theory
we propose as the dual of the deformed matrix model at tree-level. This
interpretation is consistent with our current knowledge of Liouville theory,
the checks of the string susceptibility exponents and tree-level correlation
functions in \cite{Klebanov:1994pv,Klebanov:1994kv,Barbon:1995dx} and
meshes nicely with the more general picture of double-trace
deformations in the AdS/CFT correspondence.

We propose a similar interpretation of the double scaling limit \eqref{mmabe}
that leads to the modified partition function \eqref{mmabf} of two coupled
minimal strings. In that case, we re-interpret the original product partition
sum as a function of the cosmological constant $t_-=Ct_1-St_2$ and
the dual $\tilde t_+=S\tilde t_1+C\tilde t_2$. Some evidence for this
proposal is provided in appendix \ref{app:susceptibility} where the sphere
partition sum is analyzed from the continuum formalism viewpoint.

\subsubsection{Beyond tree-level}

The full partition function of the theories we considered in
this paper is given by the Laplace transform of the original partition function.
For instance, the free energy of the standard double
scaling limit of the 2nd multi-critical matrix model is
\beq
\label{spaceca}
F(t)=-\frac{2}{5} t^{5/2}-\frac{1}{24} \log t+\frac{7}{2160} t^{-5/2}+\OO(t^{-5})
~.
\eeq
The Laplace transform of the exponential of this expression gives the
free energy of the theory modified by $(\tr \Phi^4)^2$ at the new critical point
\beq
\label{spacecb}
F(\tilde t)=\frac{3}{5} \tilde t^{5/3}-\frac{7}{36} \log \tilde t+\frac{77}{960} \tilde t^{-5/3}+
\OO(\tilde t^{-10/3})
~.
\eeq

The relation between the cosmological constant $t$ and its dual $\tilde t$ is,
up to a constant numerical factor,
\beq
\label{spacecc}
t=\tilde t^{2/3}
~.
\eeq
This relation reproduces the scaling of all terms in the perturbative expansion,
and can even be used to reproduce exactly the tree-level result with appropriate
fixing of the normalization of $t$, as explained in the previous subsection.
It fails, however, to reproduce the higher loop coefficients. For example, it fails
to reproduce the torus coefficient $-\frac{7}{36}$ giving instead the factor $-\frac{1}{36}$.

This implies that the tree-level interpretation of the previous subsection
does not extend to higher loops, where one should think of the modified
theory as a genuine NLST. It is tempting to speculate that a similar result
holds in the higher dimensional AdS/CFT example, and that also there it
is necessary to invoke the NLST structure to explain the properties of the
modified string theory in the bulk.

\subsection{Comments on the spacetime properties of the modified theories}

The overall picture for the string theory dual of two double-trace deformed
large $N$ matrix models at the critical point \eqref{mmabe}, \eqref{mmabf}
is a pair of minimal string theories with a subtle interaction induced on the level
of the worldsheet via tiny wormholes. It is interesting to ask how such interactions
manifest themselves in spacetime, or in other words how they enter in the string
field theory action.

It is apparent from the scattering amplitudes computed in section \ref{correlations}
that there will be several types of contributions to the overall effective string
field theory action, which will look schematically as
\beq
\label{sftaa}
\SS_{\rm total}=\SS_1+\SS_2+\SS_{\rm int}
~.
\eeq
$\SS_1$ $(\SS_2)$ is the effective string field theory action
of theory 1 (2). $\SS_{\rm int}$ is controlled by the double-trace angular
parameter $\theta$ (see eq.\ \eqref{mmaad}) and includes the interactions
generated by the double-trace deformation. There are interactions that involve
string fields of theory 1 or 2 separately, but also interactions coupling string
fields of different theories (the non-vanishing correlation function
$\llangle \OO_1 \OO_2 \rrangle$ in section \ref{correlations} is an example
of such interactions). $\SS_{\rm int}$ is expected to be a non-local action. 
Aspects of spacetime non-locality in NLST in the context of the AdS/CFT 
correspondence have been discussed in \cite{Aharony:2001dp,Aharony:2005sh}.

The higher dimensional generalization of this action in AdS/CFT will be
a multi-gravity theory (when the gravity approximation is justified). $1/N$ effects
will generate non-trivial interaction terms. One class of such interaction terms
includes a potential for the gravitons, in particular a mass term for a linear
combination of the gravitons. Such effective actions remind of the non-linear
bi-gravity actions of \cite{Damour:2002ws}.

\section{Discussion}
\label{discussion}

\subsection{Summary and similarities with AdS/CFT examples}

In this paper we found, by generalizing the analysis of
\cite{Klebanov:1994pv,Klebanov:1994kv}, that under certain
conditions the product of $k$ large $N$ matrix models coupled
together by a multi-trace operator admits new double scaling limits,
which define holographically $k$ $c \leq 1$ non-critical string
theories interacting with each other in a non-trivial manner.
These cases provide one- and two-dimensional illustrations of the
interacting (multi)-string theories proposed in \cite{kir,ack}.
Specific examples of this phenomenon were given in the main text for $k=2$
and in appendix \ref{app:tri} for $k=3$. At least two conditions had to
be met in order to obtain well-defined double scaling limits with a
non-vanishing multi-trace coupling: $(a)$ the single-trace operators
that participate in the deformation should share the same scaling
properties, and $(b)$ we should fine-tune the multi-trace couplings
to a special set of values. In the $k=2$ examples of section
\ref{sec:matrixmodels} this special set comprised of a one-parameter
family of double-trace couplings where the modified double scaling limit
was giving rise to a modified partition function, which is related to
the original product of two one-matrix model partition functions by
a certain Laplace transform (see eq.\ \eqref{introaa}).

We argued that these deformed large-$N$ matrix models provide
the holographic definition of an interacting product of $c\leq 1$
non-critical string theories. The interaction is mediated at the level
of the worldsheet via non-local interactions, which induce tiny wormholes
connecting worldsheets of the same or different string theories in a
manner specified explicitly by the exact matrix model free energy.
This non-local structure, which is evident in the correlation functions
presented in section \ref{correlations}, is a special example of the
non-local string theory construction anticipated on general grounds
in \cite{nonlocal} when multi-trace deformations are present in
gauge/string dualities.

During the discussion of this structure and the possible interpretations
of the modified matrix model as a modified string theory we had
the chance to revisit an old claim of Klebanov \cite{Klebanov:1994pv},
who proposed that a matrix model deformed by a critical double-trace
deformation defines a string theory with a wrong branch tachyon condensate.
We argued that one can make sense of this claim at tree-level as
a change of the bulk/boundary dictionary, or equivalently as a
re-interpretation of the standard minimal string free energy in terms
of the dual cosmological constant. This observation is analogous
to the tree-level interpretation of double-trace deformations in the
AdS/CFT correspondence as mixed boundary conditions for the dual
fields in AdS. We can check directly in the matrix model case that this
picture cannot be extended beyond tree-level, where one should
think of the modified string theories as $bona$ $fide$ NLSTs.

The matrix models in this paper and their dual string theories
are interesting because they offer a set of clean, solvable examples of
the general situation outlined in the beginning of the introduction: the
holographic duality between a product of gauge theories deformed by a
multi-trace deformation that preserves the product gauge group and string
theory on a union of spaces. In that sense, the matrix model examples
are useful precursors of analogous higher-dimensional cases in the
AdS/CFT correspondence. The latter are interesting because they give us
a rare chance to discuss the properties of massive, multi-graviton theories
in a setting with a UV completion. The higher-dimensional
AdS/CFT setups are the main focus of a companion paper \cite{kirniar2}.

Treating the matrix models as a playground for the corresponding AdS/CFT
cases is further justified by the many similarities that they exhibit. We can
summarize the basic parallels between the double-trace deformed matrix models
and their higher dimensional CFT analogs with the following three points.

\subsubsection*{Double-scaling limits vs fixed points of RG equations}

The double scaling limits in section \ref{sec:matrixmodels} zoom around points
in parameter space where the matrix models exhibit a critical behavior. At special
values of the double-trace parameters, $e.g.$ when equation \eqref{mmabd}
holds, a new critical behavior arises, but as we move away from this special
submanifold in coupling space either we recover the critical behavior of the
original undeformed  theory, or the theory enters into a branched polymer phase
where no sensible critical behavior exists. As a necessary condition for the existence
of the new critical behavior, the scaling properties of the single-trace operators
participating in the deformation have to be the same.

A similar picture of critical double-trace deformations arises in higher dimensional
QFTs \cite{kirniar2}. Let us consider a general double-trace deformation of a
$d$-dimensional CFT of the form
\beq
\label{discussaa}
\delta \SS=\int d^d x~ \Big [ g_{11}(\OO_1)^2+g_{22}(\OO_2)^2+2g_{12}\OO_1\OO_2
\Big ]~,
\eeq
where $\OO_1$, $\OO_2$ are single-trace operators in two theories (1 and 2
respectively) with scaling dimensions $\Delta_1, \Delta_2$. The deformation
\eqref{discussaa} is the direct analog of the matrix model deformation \eqref{mmaac}.
By analyzing the one-loop $\beta$-functions of the double-trace couplings in conformal
perturbation theory at leading order in $1/N$ one finds new fixed points away from the
origin when the scaling dimensions are either equal or are related by the equation
$\Delta_1=d-\Delta_2$. We encountered a similar condition on the scaling
properties of the deforming operators of the matrix models in the previous paragraph.
When new fixed points exist, there is a one-parameter family of them, which is
the analog of the one-parameter family of critical behaviors captured by the modified
free energy \eqref{introaa}. Away from the fixed points the RG flow drives us back
towards the undeformed theory at the origin, or away from the origin towards a region
where conformal perturbation theory breaks down. In the first case, the double-trace
deformation acts as an irrelevant operator around the origin -- this is the analog of
recovering the standard double scaling limit in the matrix model. The second case is
more like the branched polymer phase in the matrix model.

Beyond tree-level, the RG equations in the higher dimensional cases receive
$1/N$ corrections which displace the critical circle and introduce a coupling between
single-trace, double-trace and other multi-trace couplings. At the same time,
$1/N$ corrections in the bulk produce an effective potential for the gravitons.
In particular, they lift the mass of a linear combination of the gravitons
and make the dual multi-gravity theory massive. The structure of the RG equations
on the boundary predicts that the space will remain a union of AdS spaces only
if we fine-tune the double-trace couplings to a special set of values. Otherwise,
loop effects will backreact to the background. In those cases, where we can
trust the perturbation theory and the RG flow on the boundary drives the theory
back towards the undeformed point, we expect the dual gravity description of the IR
physics to be captured by a decoupled union of AdS spaces. In this way, the
theory develops dynamically a region of spacetime where the mass of the gravitons
is washed away. A more detailed discussion of these issues can be found in
\cite{kirniar2}.

\subsubsection*{The bulk/boundary correspondence at tree-level}

Another similarity between the matrix model and the higher dimensional
CFT cases concerns the tree-level interpretation of the deformation on the
dual string theory or gravity side. In both cases the partition
function of the deformed theory at the new critical points is related to that
of the original theory by a Legendre transform. In the bulk, we still have a
product of standard minimal string theories or a product of supergravity theories
on a union of AdS spaces, however, the bulk/boundary dictionary is now
modified. In the matrix model case, we re-interpret the standard amplitudes
of minimal strings as a function of the dual cosmological constant. In the AdS
case, we put mixed boundary conditions on the dual fields.

\subsubsection*{The NLST structure}

Finally, in both cases and to all orders in the $1/N$ expansion, the
theory in the bulk is an NLST along the lines of \cite{nonlocal}. In the
matrix model case, we have an expression exact to all orders in
perturbation theory for the modified free energy, which is given by a
Laplace transform. This allows for an explicit derivation of the NLST
rules in this case. In the higher dimensional AdS/CFT cases, we do
not have the luxury of such exact expressions but a qualitatively
similar structure is anticipated.

\subsection{Open problems and possible extensions}

We would like to close with a short list of open problems. An important
issue is the non-perturbative stability of the modified string theories in this
paper and generalizations thereof. The minimal $(p,q)$ bosonic string
theories have well-known problems at the non-perturbative level, hence
Laplace transforms of the form \eqref{mmabf}, \eqref{mmabg} are at best
good expressions in a perturbative expansion. It would be nice to re-examine
the effects of multi-trace deformations in examples which are well-defined
non-perturbatively. Perhaps, the unitary matrix models of \cite{Klebanov:2003wg}
are a good starting point for such an exercise.

In this paper we discussed exclusively what happens to closed strings when
we couple two or more non-critical string theories together. It would be interesting
to explore the effect of this coupling also to open string sectors. For instance,
FZZT branes \cite{Fateev:2000ik,Teschner:2000md} are represented in the
matrix model by the microscopic loop operator $\log \det(x-\Phi)$. It would be
interesting to analyze how the multi-trace deformations affect amplitudes that
involve FZZT branes, $i.e.$ amplitudes that involve a number of insertions of
the determinant operator $\det(x-\Phi)$, and what this implies for the worldsheet
theory.

Another interesting problem is to analyze the pattern of interactions in the case
of a large number of string theories coupled by non-local interactions. A simple
example along these lines is to consider $M$ $c=0$ matrix models coupled
pairwise with double-trace interactions in a circular fashion.
This is reminiscent of the matrix model couplings that define the $c=1$ matrix
theory in terms of $c=0$ matrix models, however, the difference here is that the
nearest-neighbor couplings are of double-trace type. Such constructions should
involve a limit where $N \gg M \gg 1$.

One of the main motivations behind this work is cosmology and the potential
for interesting time-dependent solutions in multi-gravity theories. One could
attempt to analyze such solutions in two-dimensional toy examples
extending the ideas in \cite{Karczmarek:2003pv} to systems of coupled $c=1$
string theories.

\medskip
\section*{Acknowledgements}
\noindent We would like to thank Jose Barb\'on, Igor Klebanov, David Kutasov and
Ioannis Papadimitriou for useful discussions and correspondence. We
acknowledge partial financial support by the INTAS
grant, 03-51-6346, CNRS PICS \#~2530, 3059 and 3747 and by the EU
under the contracts MEXT-CT-2003-509661, MRTN-CT-2004-503369
and MRTN-CT-2004-005104.

\appendix

 \renewcommand{\theequation}{\thesection.\arabic{equation}}
\addcontentsline{toc}{section}{Appendices}
\section*{Appendices}

\section{Products of matrix quantum mechanics models}
\label{app:MQM}

In this appendix we discuss double scaling limits in a double-trace deformation
of a product of two matrix quantum mechanics theories. Double scaling limits
in double-trace deformed matrix quantum mechanics theories were discussed
originally in \cite{Sugino:1994zr,Gubser:1994yb}.

For us the partition function of interest is\footnote{For simplicity, we will set the
ranks of the two matrix models $N_1$ and $N_2$ to be equal. This will not affect
the final results in any significant way.}
\bea
\label{appAaa}
\ZZ&=&\int D\Phi_1(t) D\Phi_2(t) ~ e^{-N \int_0^{2\pi R}dt
[\tr (\frac{1}{2} \dot\Phi_1^2+\frac{1}{2}\Phi_1^2-\lambda_1 \Phi_1^3)
+\tr (\frac{1}{2} \dot \Phi_2^2+\frac{1}{2}\Phi_2^2-\lambda_2 \Phi_2^3)]}\times
\nonumber\\
&&e^{-\int_0^{2\pi R}dt [g_{11}(\tr \Phi_1^3)^2+2g_{12} \tr \Phi_1^3 \tr \Phi_2^3+
g_{22}(\tr \Phi_2^3)^2]}
~.
\eea
The matrix model lives on a compact one-dimensional spacetime with
radius $R$. The standard double scaling limit of this model at $g_{11}=g_{22}=g_{12}=0$
describes holographically the decoupled product of two $c=1$ non-critical
string theories.

As in the zero-dimensional matrix model case of section \ref{sec:matrixmodels},
it is convenient to re-exress the double-trace deformation as
\bea
\label{appAab}
&&g_{11}(\tr \Phi_1^3)^2+g_{22}(\tr \Phi_2^3)^2+2g_{12}\tr \Phi_1^3 \tr \Phi_2^3=
\nonumber\\
&&=r_1\left( C\tr \Phi_1^3+S\tr \Phi_2^3 \right)^2
+r_2 \left( -S \tr \Phi_1^3+C\tr \Phi_2^3\right)^2
~,
\eea
where again we will use the notation $C\equiv \cos\theta$ and $S\equiv \sin \theta$.
The same relations between $g_{ij}$ and $r_1,r_2$ and $\theta$ hold as in
subsection \ref{subsec:mms}.

Following \cite{Klebanov:1994kv}, it will be useful to perform the following
Fourier transforms
\bea
\label{appAac}
P_i&&=\int_0^{2\pi R} dt~ \tr \Phi_i^3~, ~~
C_{i,n}=\sqrt 2 \int_0^{2\pi R}dt ~ \cos\frac{nt}{R} \tr \Phi_i^3~, ~ ~
\nonumber\\
S_{i,n}&&=\sqrt 2 \int_0^{2\pi R}dt ~ \sin \frac{nt}{R} \tr \Phi_i^3
~, ~ ~ i=1,2~.
\eea
Then we can recast the double-trace interactions as an infinite sum of squares
\begin{subequations}
\bea
\label{appAad}
&&\int_0^{2\pi R} dt \left( C\tr \Phi_1^3+S\tr \Phi_2^3 \right)^2=
\nonumber\\
&&=\frac{1}{2\pi R} \left\{ (CP_1+SP_2)^2 +
\sum_{n=1}^\infty \left[ (CC_{1,n}+SC_{2,n})^2+(CS_{1,n}+SS_{2,n})^2\right]
\right\}
~,
\eea
\bea
\label{appAae}
&&\int_0^{2\pi R} dt \left( -S\tr \Phi_1^3+C\tr \Phi_2^3 \right)^2=
\nonumber\\
&&=\frac{1}{2\pi R} \left\{ (-SP_1+CP_2)^2 +
\sum_{n=1}^\infty \left[ (-SC_{1,n}+CC_{2,n})^2+(-SS_{1,n}+CS_{2,n})^2\right]
\right\}
~.
\eea
\end{subequations}
and proceed following the strategy of section \ref{sec:matrixmodels}.

First, we distinguish between two cases: $\det g=r_1r_2 \neq 0$ and
$\det g=r_1r_2=0$. The former case is a simpler version of the
latter, hence this is the one we will discuss mostly. Assuming $\det g \neq 0$,
we can use repeatedly the identity \eqref{mmaaj} to recast the partition function
\eqref{appAaa} into the form
\bea
\label{appAaf}
&&\ZZ=\int_{-\infty}^\infty dy_0 dx_0 \prod_{n=1}^\infty dy_n dx_n dz_n dw_n
~e^{\frac{\pi R N^2}{2}\left[ \frac{y_0^2+y_n^2+z_n^2}{r}+
\frac{x_0^2+x_n^2+w_n^2}{\rho}\right]} \times
\nonumber\\
&&\int D\Phi_1(t) D\Phi_2(t)~ e^{-N \int_0^{2\pi R}dt
[\tr (\frac{1}{2}\dot \Phi_1^2+\frac{1}{2}\Phi_1^2-(\lambda_1+Cy_0-Sx_0)\Phi_1^3)
+\tr (\frac{1}{2}\dot \Phi_2^2+\frac{1}{2}\Phi_2^2-(\lambda_2+Sy_0+Cx_0)\Phi_2^3)]}
\times\nonumber\\
&&~~~~e^{N\sum_{n=1}^\infty \left[ (Cy_n-Sx_n)C_{1,n}+(Sy_n+Cx_n)C_{2,n}+
(Cz_n-Sw_n)S_{1,n}+(Sz_n+Cw_n)S_{2,n}\right]}
~.
\eea

In order to proceed further we need additional information about the partition
sum of a single $c=1$ MQM model. The free energy of this theory is
\begin{subequations}
\bea
\label{appAaia}
&&\log \int D\Phi(t) e^{-N\int_0^{2\pi R}dt~ \left[ \tr (\frac{1}{2}\dot \Phi^2+
\frac{1}{2}\Phi^2-(\lambda+y_0)\Phi^3\right]-\sum_{n=1}^\infty(y_n C_n+z_n S_n)}=
\nonumber\\
&&=2\pi R N^2 \left( -a_1 x+\frac{1}{2}a_2 x^2\right)+F_0(x, N^2)+F_1(x, y_n,z_n,N^2)
~,
\eea
where
\beq
\label{appAajb}
F_0(x,N^2)=RN^2\left(\frac{1}{2}a_3 x^2/\log x+\cdots\right)
-\frac{1}{24}\left( R+\frac{1}{R}\right)\log x+ \cdots
~,
\eeq
\beq
\label{appAajc}
F_1(x,y_n,z_n, N^2)=\pi R N^2 \sum_{n=1}^\infty
\left( y_n^2+z_n^2 \right) \left( b_n+{b'}_n\left(x/\log x\right)^{n/R}+
\cdots\right)+\cdots
~,
\eeq
\beq
\label{appAajd}
x=c_2-\lambda-y_0
~.
\eeq
\end{subequations}

In our case, we have to deal with the decoupled product of two such theories, so
we set
\beq
\label{appAak}
y=c_2-(\lambda_1+Cy_0-Sx_0)~, ~ ~
x=c_2-(\lambda_2+Sy_0+Cx_0)
~,
\eeq
which can be inverted to
\beq
\label{appAal}
y_0=\Delta_1-Cy-Sx~, ~ ~
x_0=\Delta_2+Sy-Cx
~.
\eeq
In these expressions $\Delta_1$ and $\Delta_2$ are still defined as in \eqref{mmaao},
\eqref{mmaaoo}, \eqref{mmaaoa}.

Plugging this information back into \eqref{appAaf} we obtain
\bea
\label{appAam}
\ZZ&=&\int_{-\infty}^\infty dxdy \prod_{n=1}^\infty dy_n dx_n dz_n dw_n~
e^{\frac{\pi R N^2}{2} \left[ \frac{y_n^2+z_n^2}{r}+\frac{x_n^2+w_n^2}{\rho}
+\frac{(\Delta_1-Cy-Sx)^2}{r}+\frac{(\Delta_2+Sy-Cx)^2}{\rho} \right]}
\times\nonumber\\
&&e^{2\pi RN^2(-a_1 x+\frac{1}{2}a_2x^2-a_1 y+\frac{1}{2}a_2y^2)+F_0(x,N^2)+F_0(y,N^2)}
\times\nonumber\\
&&e^{F_1(y,Cy_n-Sx_n,Cz_n-Sw_n,N^2)+F_1(x,Sy_n+Cx_n,Sz_n+Cw_n,N^2)}
~.
\eea
The next step is to diagonalize the quadratic terms in the auxiliary
parameters and depending on the sign of eigenvalues take a suitable
double scaling limit.

First, let us consider the quadratic part associated with the auxiliary parameters
$x,y$
\beq
\label{appAan}
\SS_2(x,y)=\frac{\pi R N^2}{2\det g} \left[ (2a_2 \det g+g_{11})x^2+
(2a_2 \det g +g_{22})y^2-2g_{12} xy \right]
~.
\eeq
To diagonalize this expression, we rotate to the variables $x_\pm$
defined by the linear transformation
\beq
\label{appAao}
y=U_y^+ x_+ +U_y^- x_-~, ~  ~ x=U^+_x x_+ + U^-_x x_-
\eeq
with $U_y^\pm$ and $U_x^\pm$ the same as $U_1^\pm$ and $U_2^\pm$
in eq.\ \eqref{mmaar}. Then we get
\beq
\label{appAap}
\SS_2(x,y)=-N^2(M_+^2 x_+^2 +M_-^2 x_-^2)
~
\eeq
with
\beq
\label{appAaq}
M^2_\pm=2\pi R m^2_\pm
\eeq
and $m^2_\pm$ as in eq.\ \eqref{mmaau}.

The corresponding masses for the higher auxiliary modes
$y_n,z_n,x_n,w_n$ are
\begin{subequations}
\beq
\label{appAwaa}
z_n,~ y_n ~ : ~ ~ m^2_{zy} = -\frac{\pi R}{2} \left(\frac{1}{r}+b_n\right)
~,
\eeq
\beq
\label{appAwab}
x_n,~ w_n ~ : ~ ~ m^2_{xw} = -\frac{\pi R}{2} \left(\frac{1}{\rho}+b_n\right)
~.
\eeq
\end{subequations}
Since $b_n<a_2$ for all $n\geq 1$ \cite{Klebanov:1994kv}, we deduce
the inequalities
\beq
\label{appAwac}
m^2_{zy}>-\frac{\pi R}{2}\left(\frac{1}{r}+2a_2\right), ~ ~
m^2_{xw}>-\frac{\pi R}{2}\left(\frac{1}{\rho}+2a_2\right)
~,
\eeq
which, say assuming $r_1>r_2$, imply
\beq
\label{appAwad}
m^2_{zy} > M_+^2~, ~ ~
m^2_{xw}> M_-^2
~.
\eeq
Consequently, the auxiliary parameters $z_n, y_n,x_n,w_n$ are massive and
can be integrated out as long as $M_\pm^2 \geq 0$. In that case, the partition
function \eqref{appAam} takes the simpler form
\begin{subequations}
\bea
\label{appAwae}
\ZZ=\int_{-\infty}^\infty dx_+ dx_- &&e^{\pi R N^2 (E^+x_+ +E^- x_-)}
e^{-N^2(m_+^2 x_+^2 +m_-^2 x_-^2)} \times
\nonumber\\
&&e^{F_0(U^+_x x_+ + U^-_x x_-, N^2)+F_0(U^+_y x_+ + U^-_y x_-, N^2)}
~,
\eea
where
\beq
\label{appAwaf}
E^{\pm}=\left(-2a_1-\frac{C\Delta_1}{r}+\frac{S\Delta_2}{\rho}\right)U^\pm_y
-\left(2a_1+\frac{S\Delta_1}{r}+\frac{C\Delta_2}{\rho}\right)U^\pm_x
~.
\eeq
\end{subequations}

If both $m_\pm^2$ are positive, we can further integrate out both $x_\pm$
to obtain the direct product of two $c=1$ partition functions
\beq
\label{appAwag}
\ZZ=e^{F_0\left(U^+_x \frac{E^+ \pi R}{2m_+^2}+U^-_x \frac{E^- \pi R}{2m_-^2},N^2\right)+
F_0\left(U^+_y \frac{E^+ \pi R}{2m_+^2}+U^-_y \frac{E^- \pi R}{2m_-^2},N^2\right)}
~.
\eeq
If one of the masses squared is zero, $m^2_+=0$, and the other positive, we can
integrate out $x_-$ to obtain a Legendre transform formula analogous
to eq.\ \eqref{mmabf}
\beq
\label{appAwai}
\ZZ=\int_{-\infty}^\infty dx_+~ e^{\pi R N^2 E^+ x_+}
e^{F_0\left( U^+_x x_+ + U^-_x \frac{E^- \pi R}{2m_-^2}, N^2\right)
+F_0\left( U^+_y x_++ U^-_y \frac{E^- \pi R}{2m_-^2}, N^2\right)}
~.
\eeq
Finally, when both $m_\pm=0$, $g_{12}=0$, $g_{11}=g_{22}=r_1=r_2=-\frac{1}{2a_2}$.
In this case, we are dealing with a decoupled product of two $c=1$
string theories that have been deformed individually by double-trace deformations
to their critical points as in \cite{Klebanov:1994kv}.

\section{Triple intersections}
\label{app:tri}

In the main text we mostly restricted our attention to
pairs of matrix models or CFTs coupled by a double-trace
deformation. In this appendix we consider a triple intersection:
the product of three theories coupled by a triple-trace operator.
As a concrete example, we will analyze the case of three 2nd multicritical
matrix models. The partition function of the theory of interest is\footnote{Again
for simplicity, we take the ranks of all three matrices to be equal.}
\beq
\label{appAar}
\ZZ=\int \prod_{i=1}^3 D\Phi_i e^{-N \sum_{i=1}^3 \left[ \frac{1}{2}\Phi_i^2+
\lambda_i \Phi_i^4 \right] -\frac{1}{N} \sum_{i,j,k=1}^3 g_{ijk}
(\tr \Phi_i^4)(\tr \Phi_j^4)(\tr \Phi_k^4)}
~.
\eeq
The coefficient of the triple-trace interaction has been scaled as $N^{-1}$,
so that the whole interaction term scales as $N^2$. By definition, the parameters
$g_{ijk}$ are fully symmetric in $i,j,k=1,2,3$.

A general multi-trace deformation that involves $n$ single-trace operators
can always be recast in terms of single-trace interactions by introducing $2n$
auxiliary field integrations \cite{Barbon:2003uu}. In the case of the
triple-trace deformation \eqref{appAar} this amounts to introducing
three auxiliary pairs of parameters $(\sigma_i,v_i)$ and writing
\bea
\label{appAas}
\ZZ&=&\int \prod_{i=1}^3 D\Phi_i d\sigma_i dv_i ~
e^{-N\sum_{i=1}^3 \left[ \frac{1}{2}\Phi_i^2 + \lambda_i \Phi_i^4 \right] -
\frac{1}{N} \sum_{i,j,k=1}^3 g_{ijk} \sigma_i \sigma_j \sigma_k +
\sum_{i=1}^3 v_i (\sigma_i - \tr \Phi_i^4)}
\nonumber\\
&=&\prod_{i=1}^3 \int d\sigma_i dx_i ~
e^{\sum_{i=1}^3 \left[ N^2 (-a_1 x_i +\frac{1}{2} a_2 x_i^2)+F(x_i,N^2)+
N(x_i-c_2-\lambda_i)\sigma_i\right]-\frac{1}{N} \sum_{i,j,k}^3 g_{ijk}
\sigma_i \sigma_j \sigma_k}
\nonumber\\
&\equiv&\prod_{i=1}^3 \int d\sigma_i d x_i~ e^{\FF(x_i,\sigma_i,N^2)}
~.
\eea
In the first equality the $v_i$ integrations are defined by analytic continuation.
In the second equality we changed integration variables from $v_i$ to $x_i$
\beq
\label{appAat}
v_i=N(x_i-c_2-\lambda_i)
~.
\eeq
The introduction of the auxiliary parameters $(\sigma_i,v_i)$ generalizes
the trick of eq.\ \eqref{mmaaj} to an arbitrary multi-trace interaction.

\subsubsection*{\it Saddle point expansion}

The saddle point equations for the integral expression \eqref{appAas} are
\begin{subequations}
\beq
\label{appAaua}
x_i~: ~ ~ -a_1+a_2 \bar x_i +\frac{1}{N^2} \frac{\d F}{\d x_i}+\frac{\bar \sigma_i}{N}=0
~,
\eeq
\beq
\label{appAaub}
\sigma_i~:~~ N(\bar x_i-c_2-\lambda_i)-\frac{3}{N} \left[
g_{iii} \bar \sigma_i^2+2 \sum_{k \neq i} g_{iik} \bar \sigma_i \bar \sigma_k
+\sum_{j,k\neq i} g_{ijk}\bar \sigma_j \bar \sigma_k \right]=0
~.
\eeq
\end{subequations}
We have denoted the saddle point values of $(\sigma_i,x_i)$ as $(\bar \sigma_i, \bar x_i)$.

The expansion of the free energy $\FF(x_i,\sigma_i,N^2)$ around the saddle point
values involves the second derivatives
\begin{subequations}
\beq
\label{appAava}
\frac{\d^2 \FF}{\d x_i \d x_j}\Bigg |_{\bar x,\bar \sigma}=
\delta_{ij} \left( N^2 a_2+\frac{\d^2 F}{\d x_i^2}\Bigg|_{\bar x}\right)
~,
\eeq
\beq
\label{appAavb}
\frac{\d^2 \FF}{\d \sigma_i \d \sigma_j}\Bigg|_{\bar x,\bar \sigma}=
-\frac{6}{N}\left[ \delta_{ij} g_{iii} \bar \sigma_i +
\left(  g_{iij} \bar \sigma_i+ g_{ijj} \bar \sigma_j + \sum_{k\neq i, k\neq j}
g_{ijk} \bar \sigma_k \right)_{i\neq j} \right]
~,
\eeq
\beq
\label{appAavc}
\frac{\d^2 \FF}{\d x_i \d \sigma_j}\Bigg|_{\bar x,\bar \sigma}=N\delta_{ij}
~.
\eeq
\end{subequations}
Several simplifications are possible in the large $N$ limit.
First, anticipating the double scaling limit, where $\bar x_i \to 0$,
we can drop the $\frac{\d^2 F}{\d x_i^2}\big|_{x_i}$ terms
in \eqref{appAava} as subleading. Then, the leading terms in the
saddle point equation \eqref{appAaua} give
\beq
\label{appAba}
\bar \sigma_i =Na_1
\eeq
and eq.\ \eqref{appAavb} becomes
\beq
\label{appAbb}
\frac{\d^2 \FF}{\d \sigma_i \d \sigma_j}\Bigg|_{\bar x,\bar \sigma}=
-6a_1 \left[ \delta_{ij} g_{iii}+\left( g_{iij}+g_{ijj}+\sum_{k\neq i, k\neq j} g_{ijk}
\right)_{i\neq j} \right]
\equiv -6 a_1 f_{ij}
~.
\eeq

With these simplifications the partition function \eqref{appAas} becomes
after shifting $x_i \to x_i -\frac{\sigma_i}{Na_2}$
\beq
\label{appAbc}
\ZZ=\prod_{i=1}^3 \int d x_i d\sigma_i~
e^{\FF(\bar x,\bar \sigma_i,N^2)+ \sum_{i=1}^3 \frac{a_2}{2}N^2 x_i^2-
\frac{1}{2} \sum_{i,j=1}^3  \MM^2_{ij} \sigma_i \sigma_j +\OO(x^3,\sigma^3)}
~,
\eeq
where we define the auxiliary field mass matrix $\MM_{ij}$ as
\beq
\label{appAbd}
\MM^2_{ij}=\frac{\delta_{ij}}{a_2}+6a_1 f_{ij}
~.
\eeq
The properties of $\ZZ$ depend crucially on whether the matrix
$\MM^2$ is positive definite or not.

As an illuminating special case, we will consider in detail
what happens when the only non-vanishing triple-trace
parameter $g_{ijk}$ is $g_{123}\equiv g$. The eigenvalues of the
mass squared matrix
\beq
\label{appAbg}
\MM^2 = \left(
\begin{array}{ccc}
\frac{1}{a_2}~ & 6a_1 g~ & 6a_1 g \\
6a_1 g & \frac{1}{a_2} & 6a_1 g \\
6a_1 g & 6a_1 g & \frac{1}{a_2}
\end{array} \right)
\eeq
are
\beq
\label{appAbi}
\lambda_- = \frac{1-6a_1a_2 g}{a_2} ~ ~ {\rm (double~ eigenvalue)}~, ~ ~
\lambda_+=\frac{1+12a_1a_2 g}{a_2}
~.
\eeq
$\MM^2$ is positive definite inside the interval
\beq
\label{appAbj}
g^*_- < g <g_+^*~, ~~ g_-^*=-\frac{1}{12a_1a_2}~, ~ g_+^*=\frac{1}{6a_1a_2}
~.
\eeq
At the lower end of this interval, $g=g_-^*$, one of the $\sigma$
eigenvectors becomes massless. At the upper end, $g=g_+^*$, two of the
$\sigma$ eigenvectors become massless. The critical values $g_\pm^*$
are therefore good candidates for the definition of a new set of double scaling
limits.

\subsubsection*{\it Double scaling limits}

We proceed to analyze the exact partition function
\bea
\label{appAbk}
\ZZ&=&\prod_{i=1}^3 \int d \sigma_i d x_i ~
e^{\sum_{i=1}^3 \left[ N^2(-a_1 x_i +\frac{1}{2}a_2 x_i^2)
+F(x_i,N^2)+N(x_i+6ga_1^2 -\Delta_i)\sigma_i\right]-
\frac{6}{N}g \sigma_1\sigma_2 \sigma_3}
\nonumber\\
&\equiv&\prod_{i=1}^3 \int d \sigma_i d x_i ~
e^{\FF(x_i,\sigma_i,N^2)}
~.
\eea
in a double scaling limit where
\beq
\label{appAbl}
N\to \infty~, ~ ~ \Delta_i \equiv c_2+\lambda_i +6ga_1^2 \to 0
\eeq
with a particular combination, to be specified, kept fixed.

It will be convenient to begin with the following manipulations:
\begin{itemize}
\item[$(i)$] Shift the auxiliary parameters $\sigma_i \to \sigma_i+N(a_1-a_2x_i)$.
\item[$(ii)$] Diagonalize the $3\times 3$ mass matrix $\MM^2$ (see eq.\ \eqref{appAbg})
with the linear transformation\footnote{The coefficients $\UU_i^j$ are functions of
$a_1,a_2,g$ whose explicit form we will not determine here.}
\beq
\label{appAbm}
x_i=\sum_j \UU^j_i X_j
~.
\eeq
\end{itemize}
In the new variables the action $\FF$ becomes
\bea
\label{appAbn}
&&\FF(\sigma_i,X_i,N^2)=\sum_{i=1}^3 \Big[ F(\UU_i^jX_j,N)-N \Delta_i \sigma_i +
N^2 a_2 \Delta_i \UU_i^jX_j - Na_2 \sigma_i \UU_i^jX_j\Big]-
\nonumber\\
&&-\frac{1}{2} N^2 a_2^2 \sum_{i,j} \Big[ \lambda_- (X_1^2+X_2^2)+\lambda_+ X^2_3\Big]
-6ga_1 \Big[ \sigma_1 \sigma_2 -Na_2(\sigma_1 \UU_2^jX_j +\sigma_2 \UU_1^jX_j)+
({\rm cyclic})\Big]-
\nonumber\\
&&-\frac{6g}{N} \Big( \sigma_1-Na_2 \UU_1^jX_j\Big)
 \Big( \sigma_2-Na_2 \UU_2^jX_j\Big) \Big( \sigma_3-Na_2 \UU_3^jX_j\Big)
~.
\eea
With minor re-arranging, we recast \eqref{appAbn} as
\bea
\label{appAbo}
\FF(\sigma_i,X_i,N^2)&=&\sum_{i=1}^3 F(\UU_i^jX_j,N)-N \Delta_i \sigma_i
-\frac{1}{2}N^2 a_2^2\left[ \lambda_-(X_1^2+X_2^2)+\lambda_+ X_3^2\right]
+G(\sigma_i)
\nonumber\\
&&+S^j(\sigma)X_j+R^{jk}(\sigma)X_jX_k+R^{ijk}X_iX_jX_k
~,
\eea
where
\begin{subequations}
\beq
\label{appAbpa}
G(\sigma)=-N\Delta_i\sigma_i-6ga_1(\sigma_1\sigma_2+
\sigma_2\sigma_3+\sigma_1\sigma_3)-\frac{6g}{N}\sigma_1\sigma_2\sigma_3
~,
\eeq
\vspace{-.7cm}
\bea
\label{appAbpb}
S^j(\sigma)&=N^2 a_2 \Delta_i \UU_i^j - Na_2 \sigma_i \UU^j_i
+6ga_1a_2N \left[ (\UU_2^j+\UU_3^j)\sigma_1+
(\UU_1^j+\UU_3^j)\sigma_2+(\UU_1^j+\UU_2^j)\sigma_3\right]
\nonumber\\
&+6ga_2\left[ \UU_3^j \sigma_1\sigma_2+ \UU_2^j \sigma_1\sigma_3
+ \UU_1^j \sigma_2\sigma_3 \right]
~,
\eea
\vspace{-.7cm}
\beq
\label{appAbpc}
R^{jk}(\sigma)=-6gNa_2^2 \left[ \UU_1^j \UU_2^k \sigma_3+
\UU_2^j \UU_3^k \sigma_1+ \UU_1^j \UU_3^k \sigma_2 \right]
~,
\eeq
\vspace{-.7cm}
\beq
\label{appAbpd}
R^{ijk}=6gN^2 a_2^3~ \UU_1^ i \UU_2^j \UU_3^k
~.
\eeq
\end{subequations}

We will consider three distinct cases: $(a)$ both $\lambda_-$, $\lambda_+>0$,
$(b)$ $\lambda_-=0$, $\lambda_+>0$ or $(c)$ $\lambda_->0$, $\lambda_+=0$.
When one of the lambdas becomes negative there will be no double-scaling
limit, so we will not discuss this situation separately.

The first case, where both lambdas are positive, occurs when $g$ lies
inside the interval \eqref{appAbj}. To cancel the linear term $X_jS^j$
in \eqref{appAbo} we shift the $X$ variables setting
\beq
\label{appAbq}
X_i=Z_i+\tilde S_i
~,
\eeq
where by definition
\beq
\label{appAbr}
\tilde S_1=\frac{S^1}{N^2a_2^2\lambda_-}~, ~ ~
\tilde S_2=\frac{S^2}{N^2a_2^2\lambda_-}~, ~ ~
\tilde S_3=\frac{S^3}{N^2a_2^2\lambda_+}
~.
\eeq
Then we scale the parameters appearing in the free energy in the following
way
\beq
\label{appAbs}
Z_i=N^{-4/5}z_i ~, ~ ~ \tilde S_i=N^{-4/5} s_i
~.
\eeq
The latter follows from the scaling
\beq
\label{appAbt}
 \Delta_i=\delta_i N^{-4/5}~, ~~ \sigma_i=N^{1/5}u_i
~.
\eeq
\eqref{appAbs} allows us to cancel the explicit dependence of the
single-matrix model free energies $F(\UU_i^jZ_j+\UU_i^j \tilde S_j, N^2)$
on $N$ and to re-express them as functions $F(\UU_i^jz_j+\UU_i^j s_j)$
of $z_j$, $s_j$ only. With this scaling the quadratic and cubic terms
$R^{jk}X_jX_k$, $R^{ijk}X_iX_jX_k$ become subleading and can be ignored.
We are left with the single-matrix free energies $F$, quadratic terms in $z_i$
and linear and quadratic terms in $s_i$. After some trivial algebra, we obtain
the following expression for the partition function
\beq
\label{appAbu}
\ZZ=\prod_{i=1}^3 \int dz_i ds_i ~
e^{\sum_{i=1}^3 F(\UU_i^j z_j+\tilde \UU_i^j s_j+t_i)}
e^{-\frac{N^{2/5}a_2^2}{2}\left[\lambda_-(z_1^2+z_2^2)+\lambda_+z_3^2\right]}
e^{-N^{2/5} d^i s_i^2}
~,
\eeq
where the new coefficients $\tilde \UU_j^i$, $d^i$ are functions of $a_1,a_2, g$
and $t_i$ are functions of $a_1,a_2,g$ and $\delta_i$ -- the scaling parameters
that were defined in \eqref{appAbt}. The last two Gaussian terms in \eqref{appAbu}
become delta functions in the large $N$ limit,\footnote{We define the large $N$
limit of the last Gaussian factor by analytic continuation when $d^i$ are negative.}
hence we end up with the partition function of three decoupled one-matrix models
\beq
\label{appAbv}
\ZZ(t_1,t_2,t_3)=\prod_{i=1}^3 e^{F(t_i)}
~.
\eeq

Now we come to the more interesting case where one of the lambda eigenvalues
vanishes. Let us consider first the case with $\lambda_-=0$,
$\lambda_+=\frac{1+12a_1a_2g}{a_2}=\frac{3}{a_2}>0$.
Eq. \eqref{appAbo} simplifies a bit to become
\bea
\label{appAbw}
\FF(\sigma_i,X_i,N^2)&=&\sum_{i=1}^3 F(\UU_i^jX_j,N)-N \Delta_i \sigma_i
-\frac{1}{2}N^2 a_2^2 \lambda_+ X_3^2 +G(\sigma_i)
\nonumber\\
&&+S^j(\sigma)X_j+R^{jk}(\sigma)X_jX_k+R^{ijk}X_iX_jX_k
~.
\eea
This time we shift the $X$ variables by setting
\beq
\label{appAca}
X_1=Z_1~, ~ ~ X_2=Z_2~, ~ ~ X_3=Z_3+\frac{S^3}{N^2 a_2^2 \lambda_+}=
Z_3+\tilde S_3
\eeq
and then we scale in the following way
\beq
\label{appAcb}
Z_i=N^{-4/5}z_i~, ~ ~ \tilde S_3=N^{-4/5}s_3~, ~ ~ \sigma_i=N^{1/5}u_i
~.
\eeq
The function $G(\sigma)$ breaks up in this limit into two pieces
scaling in different ways
\beq
\label{appAcc}
G(\sigma)=N^{2/5}\tilde G(u)-N^{6/5}\Delta_i u_i
~,
\eeq
where
\beq
\label{appAcd}
\tilde G(u)=-6ga_1(u_1u_2+u_2u_3+u_1u_3)
~.
\eeq
The last two terms in \eqref{appAbw} are again subleading in the large
$N$ limit and can be ignored. Hence, we are left at this stage with
the action
\bea
\label{appAce}
\FF(u_i,z_i,N^2)&=&\sum_{i=1}^3 F(\UU_i^j z_j+\UU_i^3 s_3)
+N^{2/5} \tilde G(u)-N^{6/5} \Delta_i u_i
\nonumber\\
&&+N^{-4/5}(z_1S^1+z_2S^2)-\frac{1}{2}N^{2/5}a_2^2\lambda_+z_3^2+
\frac{1}{2}N^{2/5}a_2^2\lambda_+ s_3^2
~.
\eea
The $z_3^2$ term in \eqref{appAce} will contribute a delta function and
the $z_3$ integral will localize at $z_3=0$. This simplifies the action $\FF$
a bit further to
\bea
\label{appAcf}
\FF(u_i,z_1,z_2,N^2)&=&\sum_{i=1}^3 F(\UU_i^1 z_1+\UU_i^2z_2+\UU_i^3 s_3)
+N^{2/5} \tilde G(u)-N^{6/5} \Delta_i u_i
\nonumber\\
&&+N^{-4/5}(z_1S^1+z_2S^2)+\frac{1}{2}N^{2/5}a_2^2\lambda_+ s_3^2
~.
\eea

The term $N^{-4/5}(z_1S^1+z_2S^2)$ has a $u_i$-independent piece
\beq
\label{appAcg}
N^{6/5}a_2 \Delta_i (\UU^1_iz_1 +\UU^2_i z_2)
\eeq
which we will require to stay finite in the large $N$ limit. This can
be achieved with the scaling
\beq
\label{appAci}
\UU_i^\alpha\Delta_i=N^{-6/5}\delta^\alpha~, ~ ~ \alpha=1,2
~.
\eeq
The remaining combination of $\Delta_i$'s will be scaled in the standard
way
\beq
\label{appAcia}
\UU^3_i \Delta_i=N^{-4/5}\delta
~.
\eeq
Hence, after some algebra $\FF$ reduces to an expression of the form
\bea
\label{appAcj}
\FF(u_i,z_1,z_2,N^2)&=&\sum_{i=1}^3 F(\UU_i^1 z_1+\UU_i^2z_2
+\UU_i^3\frac{\delta}{a_2\lambda_+}+\UU_i^3 \NN^j u_j)
+a_2 \delta_i(\UU_i^1z_1+\UU_i^2z_2)
\nonumber\\
&&+N^{2/5}\left[ u_i(\KK^{1i}z_1+\KK^{2i}z_2)+\LL^{ij}u_iu_j \right]+
\PP^i u_i
~,
\eea
where the constants $\NN^j,\KK^{\alpha i},\LL^{ij},\PP^i$ (for $i,j=1,2,3$,
$\alpha=1,2$) depend only $a_1,a_2$.

To obtain the final result, we diagonalize the
quadratic term $\LL^{ij}u_i u_j$ and use the large $N$ limit to localize
the $u_i$ integrals at $u_i=0$. This kills all the contributions which are linear
and homogeneous in $u_i$. Then, defining the
new scaling parameters
\beq
\label{appAck}
\tilde t^1=a_2 \delta_i \UU_i^1~, ~ ~ \tilde t^2=a_2 \delta_i \UU_i^2~, ~ ~
t_3=\frac{\delta}{a_2\lambda_+}
\eeq
and renaming for aesthetic reasons $z_1$, $z_2$ as $t_1, t_2$ we get
the double-scaled partition function
\beq
\label{appAcl}
\ZZ(\tilde t_1,\tilde t_2,t_3)=\int_{-\infty}^\infty dt_1 dt_2~
e^{\tilde t^1 t_1+\tilde t^2 t_2+\sum_{i=1}^3 F(\UU_i^1 t_1+\UU_i^2 t_2+\UU_i^3 t_3)}
~.
\eeq
This expression, which generalizes eq.\ \eqref{mmabf}, is a double Laplace transform
of the original partition function. A double integration is natural, since it is a
double eigenvalue ($i.e.$ $\lambda_-$) that we tune to zero.

Similar manipulations can be performed in the last case of interest:
$\lambda_-=\frac{3}{2a_2}>0$, $\lambda_+=0$. We will skip
the gory details and present the final, double-scaled expression
for the partition function
\beq
\label{appAcl}
\ZZ(\tilde t_1, t_2,t_3)=\int_{-\infty}^\infty dt_1~
e^{\tilde t^1 t_1+\sum_{i=1}^3 F(\UU_i^1 t_1+\UU_i^2 t_2+\UU_i^3 t_3)}
~.
\eeq

\section{String susceptibility exponents in the continuum approach}
\label{app:susceptibility}

The double scaling limit \eqref{mmabe} and the resulting partition
function \eqref{mmabf} give the holographic definition of two interacting
minimal (2,3) bosonic string theories. In this appendix, we present evidence
for the tree-level interpretation of these theories in subsection
\ref{treelevel} by studying the scaling properties of the deformed sphere
partition function using the continuum formalism of the minimal strings.

For completeness, and in order to set the notation, let us begin by recalling
a few well-known facts about the one-matrix model case. For the main example
in this appendix, the double scaling limit of the partition function of the 2nd
multicritical matrix model
\beq
\label{wsaa}
\ZZ=\int D\Phi~ e^{-N\tr\left[\frac{1}{2} \Phi^2+\lambda \Phi^4\right]}
~
\eeq
one finds the free energy
\beq
\label{wsab}
F(t)=-\frac{2}{5} t^{5/2}-\frac{1}{24} \log t+\frac{7}{2160} t^{-5/2}+\OO(t^{-10})
~
\eeq
where $t$ is the double scaling parameter.

When we add a double-trace deformation to obtain the partition function
\beq
\label{wsac}
\ZZ=\int D\Phi~ e^{-N\tr \left[ \frac{1}{2}\Phi^2+\lambda \Phi^4\right] - g (\tr \Phi^4)^2}
\eeq
and tune $g$ to the critical value $-\frac{1}{2a_2}$, a new double-scaling
limit is possible giving the free energy
\beq
\label{wsad}
F(\tilde t)=\frac{3}{5} \tilde t^{5/3}-\frac{7}{36} \log \tilde t+\frac{77}{960} \tilde t^{-5/3}
+ \cdots
~.
\eeq
This expression can be deduced from \eqref{wsab} and the Laplace transform
\beq
\label{wsae}
F(\tilde t)=\log \int_{-\infty}^\infty dt~ e^{t \tilde t+F(t)}
~.
\eeq

The string susceptibility exponents $\gamma$ are defined at any genus $g$
in terms of the exponents of the scaling parameters at each order in the expansion
of the free energy
\beq
\label{wsaf}
F(t)=...+\#~ t^{(2-\gamma)\chi/2}+...~, ~ ~ \chi=2-2g
~.
\eeq
The above double scaling limits exhibit different string susceptibility exponents:
\eqref{wsab} gives $\gamma=-\frac{1}{2}$, whereas
\eqref{wsad} gives $\gamma=\frac{1}{3}$.

There is a simple way to determine the critical exponents $\gamma$ from the
continuum formulation of the dual string theory. For $(p,q)$ minimal models coupled
to gravity the Liouville interaction takes the form
\beq
\label{wsaga}
\delta \SS_{Liouville}=\mu \int d^2 z ~ \OO_{min} ~e^{\alpha_+ \phi}
~, ~ ~ ~
\alpha_+=-\frac{p+q-1}{\sqrt{2pq}}
\eeq
where $\OO_{min}$ is the matter primary field with the lowest dimension
\beq
\label{wsagb}
h_{min}=\frac{1-(p-q)^2}{4pq}
~.
\eeq
The Liouville path integral in the partition function can be computed by separating
the zero mode $\phi_0$ and performing the relevant integral
\beq
\label{wsai}
\int_{-\infty}^\infty d\phi_0 ~ e^{Q\chi \phi_0/2- \mu ~e^{\alpha_+ \phi_0}}=
\frac{1}{\alpha_+} \Gamma\left( \frac{Q \chi}{2\alpha_+} \right)
\mu^{-\frac{Q \chi}{2\alpha_+}}
~.
\eeq
In this expression $Q=\sqrt 2 \frac{p+q}{\sqrt{pq}}$ is the linear dilaton slope.
Identifying the Liouville interaction constant $\mu$ with the matrix model
scaling parameter $t$ we deduce the critical exponent
\beq
\label{wsaj}
\gamma=2+\frac{Q}{\alpha_+}
~.
\eeq
For example, when $(p,q)=(2,3)$ (corresponding to the 2nd multicritical matrix model)
we have $Q=\frac{5}{\sqrt 3}$ and $\alpha_+=-\frac{2}{\sqrt 3}$ and therefore
\beq
\label{wsak}
\gamma=-\frac{2}{p+q-1}=-\frac{1}{2}
~
\eeq
reproducing the above matrix model result.

For the other double scaling limit \eqref{wsad}, which involves a critical
double-trace deformation, it has been proposed by ref.\ \cite{Klebanov:1994pv}
that one should consider a (2,3) minimal string with the wrong branch tachyon
in \eqref{wsaga}, $i.e.$ one should replace
$\alpha_+ \to \alpha_-=-\frac{p+q+1}{\sqrt{2pq}}$. Indeed, for $(p,q)=(2,3)$
this substitution reproduces the matrix model result $\gamma=1/3$.
In subsection \ref{treelevel} we rephrased this proposal as a
transformation from the cosmological constant $\mu$ to the dual
cosmological constant $\tilde \mu$.

We now proceed to apply a similar logic to the two-matrix model
case \eqref{mmabf}. The sphere contribution to the free energy
\beq
\label{wsal}
F(\tilde t_+,t_-)=\log \int_{-\infty}^\infty dt_+~
e^{\tilde t_+ t_+ +F_1(U_1^+ t_+ +U_1^- t_-)+F_2(U_2^+ t_+ +U_2^- t_-)}
\eeq
is the leading contribution in the saddle point approximation.
The saddle point value of $t_+$ is given implicitly by the following equation
\beq
\label{wsam}
\tilde t_+=U^+_1 (U_1^+ t_+ +U_1^- t_-)^{3/2}+
U_2^+ (U_2^+ t_+ +U_2^- t_-)^{3/2}
~.
\eeq
We should solve this equation for $t_+$ in terms of $\tilde t_+, t_-$
and then insert the result into the saddle point expression for $F$
\beq
\label{wsan}
F(\tilde t_+,t_-)=\tilde t_+ t_+ -\frac{2}{5} (U_1^+ t_+ +U_1^- t_-)^{5/2}-
\frac{2}{5} (U_2^+ t_+ +U_2^- t_-)^{5/2}+ \cdots
~.
\eeq

It seems difficult to obtain a closed expression for generic $t_-$, but one
can easily deduce an expression that involves a perturbative expansion in $t_-$.
Indeed, one can show that the sphere contribution to the free energy admits
an expansion of the form
\beq
\label{wsao}
F(\tilde t_+,t_-)=\sum_{n=0}^\infty f_n(\theta)~ \tilde t_+^{\frac{5-2n}{3}} t_-^n+ \cdots
~,
\eeq
where $f_n(\theta)$ are functions of $\theta$ that can be determined.
In an effort to reproduce this expansion from Liouville theory, we now turn
to the continuum formalism of the minimal string.

Before adding the double-trace deformation, the total free energy
$F(t_1,t_2)$ is simply the sum of the free energies of the two independent
constituent matrix models and the sphere contribution has the following
form in terms of Liouville zero mode integrals
\beq
\label{wsap}
F(t_1,t_2)\big|_{\rm sphere}\sim \int_{-\infty}^\infty d\phi_1 e^{Q\phi_1-t_1 e^{\alpha_+ \phi_1}}
+ \int_{-\infty}^\infty d\phi_2 e^{Q\phi_2-t_2 e^{\alpha_+ \phi_2}}
\sim t_1^{-\frac{Q}{\alpha_+}} +t_2^{-\frac{Q}{\alpha_+}}= t_1^{5/2}+t_2^{5/2}
~.
\eeq
Rotating to a new mixed basis of coupling constants
\beq
\label{wsaq}
t_1=U_1^+ t_+ + U_1^- t_-~, ~ ~ t_2=U_2^+ t_+ +U_2^- t_-
\eeq
we may rewrite the same expression as
\beq
\label{wsar}
F(t_+,t_-)\big|_{\rm sphere} \sim \int_{-\infty}^\infty d\phi_1 e^{Q\phi_1-
(U_1^+t_++U_1^- t_-) e^{\alpha_+ \phi_1}}
+ \int_{-\infty}^\infty d\phi_2 e^{Q\phi_2-
(U_2^+t_+ +U_2^- t_-) e^{\alpha_+ \phi_2}}
~.
\eeq

To obtain the worldsheet version of the matrix model result \eqref{wsal},
\eqref{wsao} we transform from $t_+e^{\alpha_+\phi_{1,2}}$ to the dual
cosmological constant interactions $\tilde t_+e^{\alpha_- \phi_{1,2}}$
leaving the $t_-$ terms invariant. This implies the Liouville interaction
\bea
\label{wsas}
\delta \SS_{{\rm total}}=
\int d^2 z_1 \left( U_1^+ \tilde t_+ e^{\alpha_- \phi_1}+ U_1^- t_- e^{\alpha_+ \phi_1} \right)
+\int d^2 z_2 \left( U_2^+ \tilde t_+ e^{\alpha_- \phi_2}+ U_2^- t_- e^{\alpha_+ \phi_2} \right)
~.\nonumber\\
\eea
The sphere free energy is then
\bea
\label{wsat}
F(\tilde t_+,t_-)\sim
\int_{-\infty}^\infty d\phi_1
e^{Q\phi_1- U_1^+ \tilde t_+ e^{\alpha_-\phi_1} - U_1^- t_- e^{\alpha_+ \phi_1}}
+\int_{-\infty}^\infty d\phi_2
e^{Q\phi_2- U_2^+ \tilde t_+ e^{\alpha_-\phi_2} - U_2^- t_- e^{\alpha_+ \phi_2}}
~.\nonumber\\
\eea
Expanding the exponentials in powers of $t_-$ and then evaluating
the Liouville zero mode integrals term by term we obtain
\bea
\label{wsau}
F(\tilde t_+,t_-)\big|_{\rm sphere}\sim \sum_{n=0}^\infty h_n(\theta)
~ \tilde t_+ ^{-\frac{Q+n\alpha_+}{\alpha_-}} ~ t_-^n=
\sum_{n=0}^\infty h_n(\theta)~ \tilde t_+^{\frac{5-2n}{3}} t_-^n
\eea
reproducing the matrix model expansion \eqref{wsao}. We have not attempted
to make a precise matching of the functions $f_n(\theta), h_n(\theta)$.
This would involve going beyond the zero mode integrals. We regard
the expansion \eqref{wsau} as evidence in favor of the picture proposed
in subsection \ref{treelevel}.


\end{document}